\documentclass{aa}
\usepackage{txfonts}
\usepackage{epsfig}
\usepackage{subfigure}
\usepackage{multirow}

\usepackage{times}
\usepackage{natbib}
\bibpunct{(}{)}{;}{a}{}{,}

\begin{document}

\title{The soft X--ray/NLR connection: a single photoionized medium?}
\author{Stefano Bianchi\inst{1}, Matteo Guainazzi\inst{1}, Marco Chiaberge\inst{2,3}}

\offprints{Stefano Bianchi\\ \email{Stefano.Bianchi@sciops.esa.int}}

\institute{XMM-Newton Science Operations Center, European Space Astronomy Center, ESA, Apartado 50727, E-28080 Madrid, Spain
\and Space Telescope Science Institute, 3700 San Martin Drive, Baltimore, MD 21218
\and INAF-Istituto di Radioastronomia, Via P. Gobetti 101, 40129 Bologna, Italy}

\date{Received / Accepted}

\authorrunning{S. Bianchi et al.}

\abstract{We present a sample of 8 nearby Seyfert 2 galaxies observed by \textit{HST} and \textit{Chandra}. All of the sources present soft X-ray emission which is coincident in extension and overall morphology with the [{O\,\textsc{iii}}] emission. The spectral analysis reveals that the soft X-ray emission of all the objects is likely to be dominated by a photoionized gas. This is strongly supported by the 190 ks combined XMM-\textit{Newton}/RGS spectrum of Mrk~3, which different diagnostic tools confirm as being produced in a gas in photoionization equilibrium with an important contribution from resonant scattering. We tested with the code \textsc{cloudy} a simple scenario where the same gas photoionized by the nuclear continuum produces both the soft X-ray and the [{O\,\textsc{iii}}] emission. Solutions satisfying the observed ratio between the two components exist, and require the density to decrease with radius roughly like $r^{-2}$, similarly to what often found for the Narrow Line Region.

\keywords{galaxies: Seyfert - X-rays: galaxies}

}

\maketitle

\section{Introduction}

Highly obscured objects are ideal laboratories to study the properties of the gas in the environment of Active Galactic Nuclei (AGN). The high energy X-ray spectrum is dominated by a heavily absorbed power law and, especially when the column density is very large (as for the Compton-thick objects, with $N_H>\sigma_t^{-1}\simeq1.5\times10^{24}$ cm$^{-2}$), by Compton reflection and a strong iron fluorescence line. Both the last two components are likely to be produced in a compact (parsec-scale), Compton-thick and low ionized material, traditionally identified with the absorber along the line of sight \citep[the so-called `torus' envisaged in the Seyfert unification scenarios:][]{antonucci93}.

On the other hand, the soft X-ray spectrum had to await the high resolution spectrometers aboard XMM-\textit{Newton} and \textit{Chandra} to reveal its real nature, at least in the three brightest sources, where such an experiment is possible: NGC~1068 \citep{kin02,brink02,ogle03}, Circinus \citep{Sambruna01b} and Mrk~3 \citep{sako00b,bianchi05b,pp05}. In these cases, the spectrum is dominated by emission lines, mainly from He-- and H--like K transitions of light metals and L transitions of Fe. The `continuum' observed in lower resolution CCD spectra was mainly due to the blending of these features. Different diagnostic tests agree that in all these sources the observed lines are more easily explained if produced in a gas photoionized by the AGN, rather than in a hot gas in collisional equilibrium.

The unprecedented high spatial resolution of \textit{Chandra} added another key ingredient to the puzzle. The same three sources revealed that their soft X-ray emission was actually extended on hundreds of pc, thus being clearly produced in a gas well beyond the torus \citep{yws01,Sambruna01a,sako00b}. Moreover, the dimension and morphology of this emission closely resemble that of the Narrow Line Region (NLR), as mapped by the [{O\,\textsc{iii}}] $\lambda5007$ emission line. Since the NLR is also generally believed to be mainly constituted of gas photoionized by the nuclear continuum, this correlation deserves a deeper investigation. Some positive results in this direction have already been published on single sources \citep[see e.g.][]{kin02,ogle03}.

We present in this paper a sample of 8 Seyfert 2 galaxies with extended [{O\,\textsc{iii}}] emission in their \textit{HST} images, which were also observed by \textit{Chandra} (Sect. \ref{obs}). We show in Sect. \ref{imaging} that all of them present a soft X-ray emission with extension and morphology highly correlated with that of the NLR. Moreover, the spectral analysis generally suggests that the most likely origin for this emission is from a photoionized gas (Sect. \ref{spectral}). Therefore, we produce photoionization models to find out if the same gas can be actually responsible for both components (Sect. \ref{cloudy}), and finally discuss on the results (Sect. \ref{discussion}).\\

The cosmological parameters used throughout this paper are $H_0=70$ km s$^{-1}$ Mpc$^{-1}$, $\Lambda_0=0.73$ and $q_0=0$ \citep[i.e. the default ones in \textsc{Xspec} 12.2.0:][]{xspec}.

\section{\label{obs}Observations}

\subsection{The sample}

Table \ref{log} presents the sources included in the sample under analysis, along with some basic properties and the details on the \textit{Chandra} and \textit{HST} observations analyzed in this paper. The sample consists of all the Seyfert 2 galaxies included in the \citet{schm03} catalog, with a \textit{Chandra} observation. The complete \citet{schm03} catalog is selected on the basis of 60 $\mu$m luminosities and warm 25/60 $\mu$m colors, in order to avoid starburst galaxies and orientation biases, and includes only Seyfert galaxies with $z<0.031$. We excluded NGC~1068, because its \textit{Chandra} image and correlation with optical and radio data were treated in detail by \citet{yws01}. On the other hand, we added another source, NGC~5643, whose [{O\,\textsc{iii}}] image was discussed by \citet{sim97}.

\begin{table*}
\caption{\label{log}The log of all the observations analysed in this paper.}

\begin{center}
\begin{tabular}{ccccccccccccc}
\hline
\textbf{Name} 	& \textbf{z} 	& \textbf{1$\arcsec$}$^a$	&\multicolumn{3}{c}{\textbf{\textit{Chandra}}}	& \multicolumn{3}{c}{\textbf{\textit{HST}}} &\textbf{F}$_\mathrm{0.5-2\,keV}$ $^f$ & \textbf{F}$_\mathrm{[{O\,\textsc{iii}}]}$ $^g$& \textbf{[{O\,\textsc{iii}}]/}	\\
		&		& \textbf{(pc)}		& \textbf{Date}	& \textbf{Instr.}$^b$ &\textbf{Exp.$^c$}	& \textbf{Prog. ID}	& \textbf{I/F}$^d$ &\textbf{Exp.$^e$} & ($10^{-13}$ \textbf{cgs}) & ($10^{-13}$ \textbf{cgs}) &	\textbf{soft X}\\
\hline
&&&&&&&&&&&\\
NGC~1386	& 0.0029 	&	60		& 2003-11-19 & A-S	& 20 			& 6419 & W2/F5 & 800 & 1.8 & 5.1 & 2.8\\
Mrk~3  		& 0.0135 	&	275		& 2000-03-18 & A-S/H	& 100			& 5140 & F/F5 & 750 & 4.7 & 20.5 & 4.4\\
NGC~3393  	& 0.0125 	&	255		& 2004-02-28 & A-S	& 29			& 3981 & W/F5 & 8000 & 2.2 & 10.5 & 4.8\\
NGC~4388	& 0.0084	&	170		& 2001-06-08 & A-S	& 20			& 6332 & W2/FR & 280 & 3.4 & 13.4 & 3.9\\
NGC~4507	& 0.0118	&	240		& 2001-03-15 & A-S/H	& 140			& 8259 & W2/FR & 520 & 3.3 & 11.0 & 3.3\\
NGC~5347  	& 0.0078 	&	160		& 2004-06-05 & A-S	& 37			& 8598 & W2/FR & 800 & 0.27 & 0.8 & 3.0\\
NGC~5643  	& 0.0040 	&	80		& 2004-12-26 & A-S	& 8			& 5411 & W2/F5 & 700 & 1.4 & 5.1 & 3.6 \\
NGC~7212  	& 0.0266 	&	530		& 2003-07-22 & A-S	& 20			& 6332 & W2/FR & 600 & 0.8 & 8.8 & 11.0\\
&&&&&&&&&&&\\
\hline
\end{tabular}
\end{center}

$^a$ Radial extent of 1 arcsec at the distance of the source, assuming the cosmology reported in the text

$^b$ \textit{Chandra} instrument (A-S: ACIS-S; A-S/H: ACIS-S/HETG)

$^c$ \textit{Chandra} exposure time in ks

$^d$ \textit{HST} instrument/filter configuration (W: WFPC; W2: WFPC2; F: FOC; F5: F502N; FR: FR533N)

$^e$ \textit{HST} exposure time in s

$^f$ All values taken from this paper, except for NGC~4388 \citep{iwa03}

$^g$ All values taken from \citet{schm03}, except for NGC~5643 (this paper)
\end{table*}

\subsection{\label{datared}Data reduction}

\subsubsection{\label{chandra}\textit{Chandra}}

We report here for the first time on the \textit{Chandra} ACIS observations of all the sources in the sample, with the exceptions of Mrk~3 \citep{sako00b}, NGC~4388 \citep{iwa03} and NGC~4507 \citep{matt04b}. All data were reduced with the Chandra Interactive Analysis of Observations (\textsc{ciao}) 3.2.1 and the Chandra Calibration Database (\textsc{caldb}) 3.0.3, adopting standard procedures. In particular, a new evt2 file was created with \texttt{acis\_process\_events}, adopting an observation-specific bad pixel file and dedicated background-cleaning procedures in the cases of \textsc{vfaint} observations. Moreover, filtering of periods of anomalous background levels and correction for ACIS contamination at short wavelengths were performed. Spectra were extracted with \textsc{psextract}, adopting three types of regions for each source: 1) circular regions with radius of 1 arcsec (the `nucleus' hereafter); 2) annuli with inner radius of 1 arcsec and a different outer radius for each source, depending on the size of the extended region (`extended region' hereafter); 3) circular regions encompassing both. All images were corrected for any known offset, to achieve the best possible accuracy in astrometry.

All spectra were binned in order to have no less than 25 counts in each spectral channel, allowing us to use the $\chi^2$ statistics. On the other hand, `local fits' were also performed in the 5.25-7.25 keV energy range with the unbinned spectra, using the \citet{cash76} statistics, in order to better assess the presence of the iron lines and measure their properties. We refer the reader to \citet{gua05b} for details on this kind of analysis.

In the following, we will always call `soft' and `hard' the X-ray spectra below and above 3 keV, respectively. Fit intervals depend on the quality of the spectra, but are in any case restricted to the range 0.3-9.5 keV, which is approximately where the ACIS CCDs are well calibrated.

\subsubsection{\label{hst}\textit{HST}}

We downloaded the narrowband optical images from the MAST (multi--mission archive at STScI). The observations were made as part of different programs, and with different instruments, filters and exposure  times. The log of \textit{HST} observations is reported in Table \ref{log}. The images were processed through the standard OTFR (on-the-fly reprocessing) calibration pipeline which performs analog-to-digital conversion, bad pixel masking, bias and dark subtraction, flat field correction and photometric calibration. The data, usually a set of two images for each object, are then combined to reject cosmic rays. The images taken with the FOC are corrected for geometric distorsion by the pipeline, while for WFPC2 data we used the \texttt{multidrizzle} script \citep{koe02}.

A crucial issue is clearly represented by the relative astrometry of the images from \textit{Chandra} and \textit{HST}. \textit{Chandra} has a nominal position accuracy of 0.6 arcsec (at the 90\% confidence level), while the absolute astrometry of \textit{HST} is accurate to 1-2 arcsec. Since the NLR are extended on scales of few arcsec, the systematic offset between the images can be relevant. Unfortunately, in none of the datasets included in our sample point-like sources are present in the fields of both instruments that could help to align the two astrometries. We therefore decided to match the two pixels with the largest count-rate, which should represent the nucleus in both wavebands. The uncertainty relating to the fact that these sources are obscured, which, in principle, would displace the optical from the X-ray peak, should not be very relevant, because the physical size of the \textit{Chandra} pixel (around 0.5 arcsec) is much larger than the scale where the obscuration takes place (see Table \ref{log}). The reliability of this method, at least for the aims of this paper, is supported by the case of NGC~1068, whose \textit{HST} and \textit{Chandra} peaks, independently aligned, are coincident within the \textit{Chandra} errors \citep{yws01}. Mrk~3 represents the only exception, since, as already noted by e.g. \citet{ruiz05}, it lacks a bright central [{O\,\textsc{iii}}] knot. In this case, we assumed that the optical nucleus were in the center of symmetry of the observed structure, which also coincides with the pixel with the lowest count-rate.

\subsubsection{XMM-\textit{Newton}}

In this paper we present the high-resolution combined spectrum of Mrk~3, because this is the only source in the sample whose RGS spectrum has a quality good enough to provide us with useful gas diagnostics constraints. Mrk~3 was observed 10 times by XMM-\textit{Newton} between 2000 and 2002. In all these observations, the Radiation Grating Spectrometer \citep[RGS;][]{denher_rgs01} was operating in standard spectroscopic mode. Part of the data of these observations were discussed by \citet{bianchi05b} and \citet{pp05}. We have reduced the data of all the RGS XMM-\textit{Newton} observations of Mrk~3, using the standard SAS \citep{sas610} meta-task \texttt{rgsproc}, and the most updated calibration files available at the moment the reduction was performed (July 2005). Each spectrum was produced assuming that the X-ray emission comes from a point source located at the optical coordinated of the Mrk~3 AGN, thus ensuring an homogeneous wavelength reconstruction. The wavelength systematic uncertainty is ~8 m$\AA$ across the whole RGS sensitive bandpass. Spectral fits have been performed on two 1st order RGS1 and RGS2 spectra, resulting from the sum of all the available spectra, generated by \texttt{rgscombine}. Fig. \ref{mrk3rgs} - which we present for illustrative purpose only in Sect. \ref{mrk3} - was generated by combining all the 1st and 2nd order spectra of the two RGS cameras, and smoothing the combined spectra through a convolution with a 5-spectral channel wide triangular function.

All spectra were analyzed with \textsc{Xspec} 12.2.0 \citep{xspec}. In the following, errors correspond to the 90\% confidence level for one interesting parameter ($\Delta \chi^2 =2.71$), where not otherwise stated.

\section{\label{analysis}Analysis}

\subsection{\label{imaging}Imaging}

In none of the sources there is evidence of hard X-ray emission extended beyond the instrument Point Spread Function (PSF), in agreement with an origin of this component within a few pc from the nucleus. On the other hand, most of the objects present a clear extension in the soft X-ray emission, with 50 to 80\% of soft X-ray counts detected outside the inner 1 arcsec (which corresponds to 60-530 pc, depending on the source: see third column of Table \ref{log}). These percentages are much higher than those expected for an on-axis point-like source observed by ACIS (see Fig. 4.7 and Table 4.2 of The Chandra Proposers' Observatory Guide\footnote{http://cxc.harvard.edu/proposer/POG/html/MPOG.html}). The soft X-ray emission is much more concentrated  in NGC~7212, where only 25\% of the counts are detected outside the inner 1 arcsec. This is not surprising, as this source is by far the farthest of our sample (see Table \ref{log}). However, the emission is strongly elongated and so clearly extended. Finally, \citet{matt04b} report no extension in the \textit{Chandra} image of NGC~4507. We also do not find any conclusive evidence in favour of an extended soft X-ray emission in our re-analysis of the data. However, it should be noted that the observation is severely affected by pile-up (see Sect. \ref{datared}), which prevents us from performing a quantitative comparison with the PSF of the instrument.

\begin{figure*}

\begin{center}
\epsfig{file=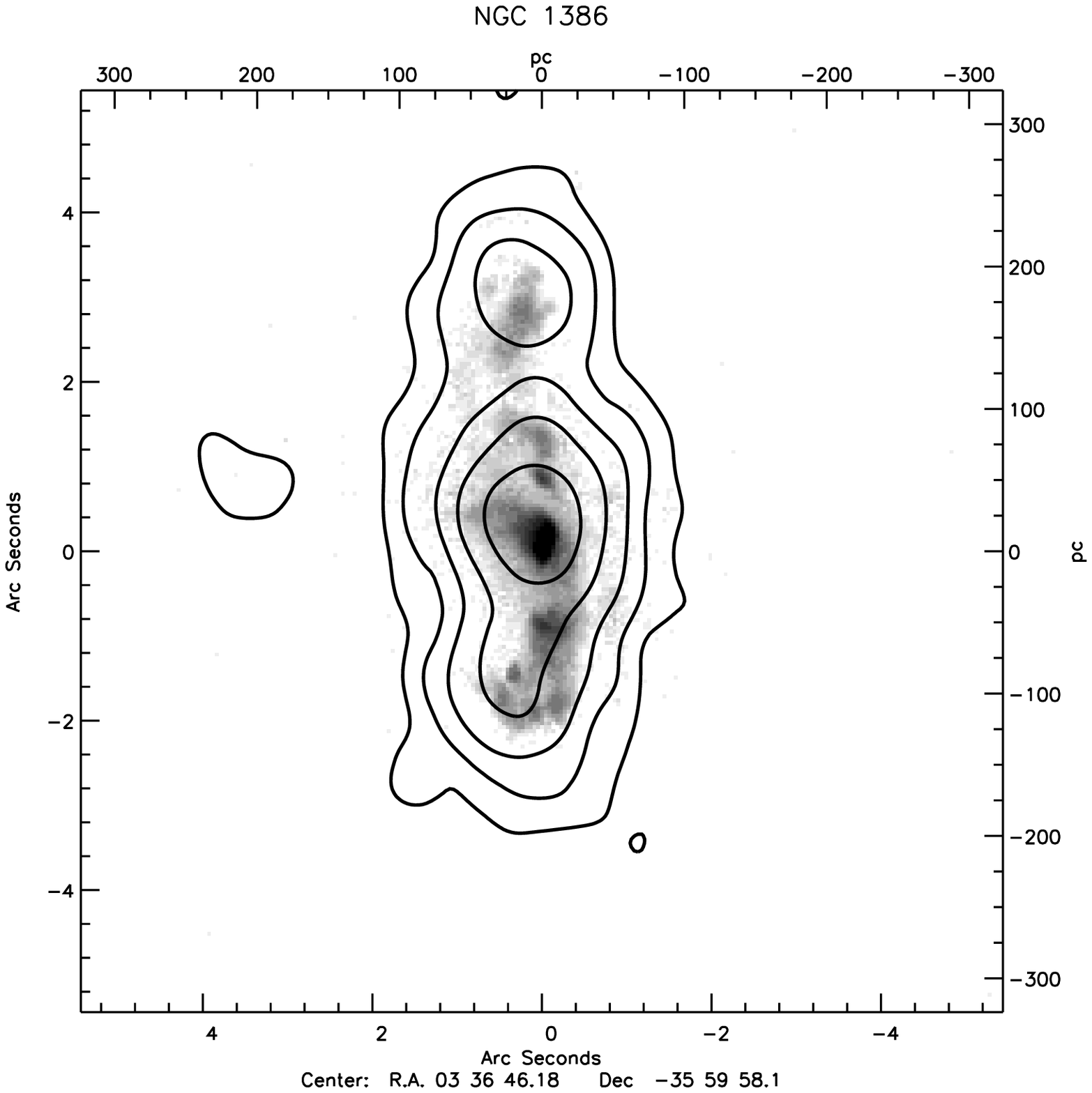, width=6.92cm}
\hspace{0.5cm}
\epsfig{file=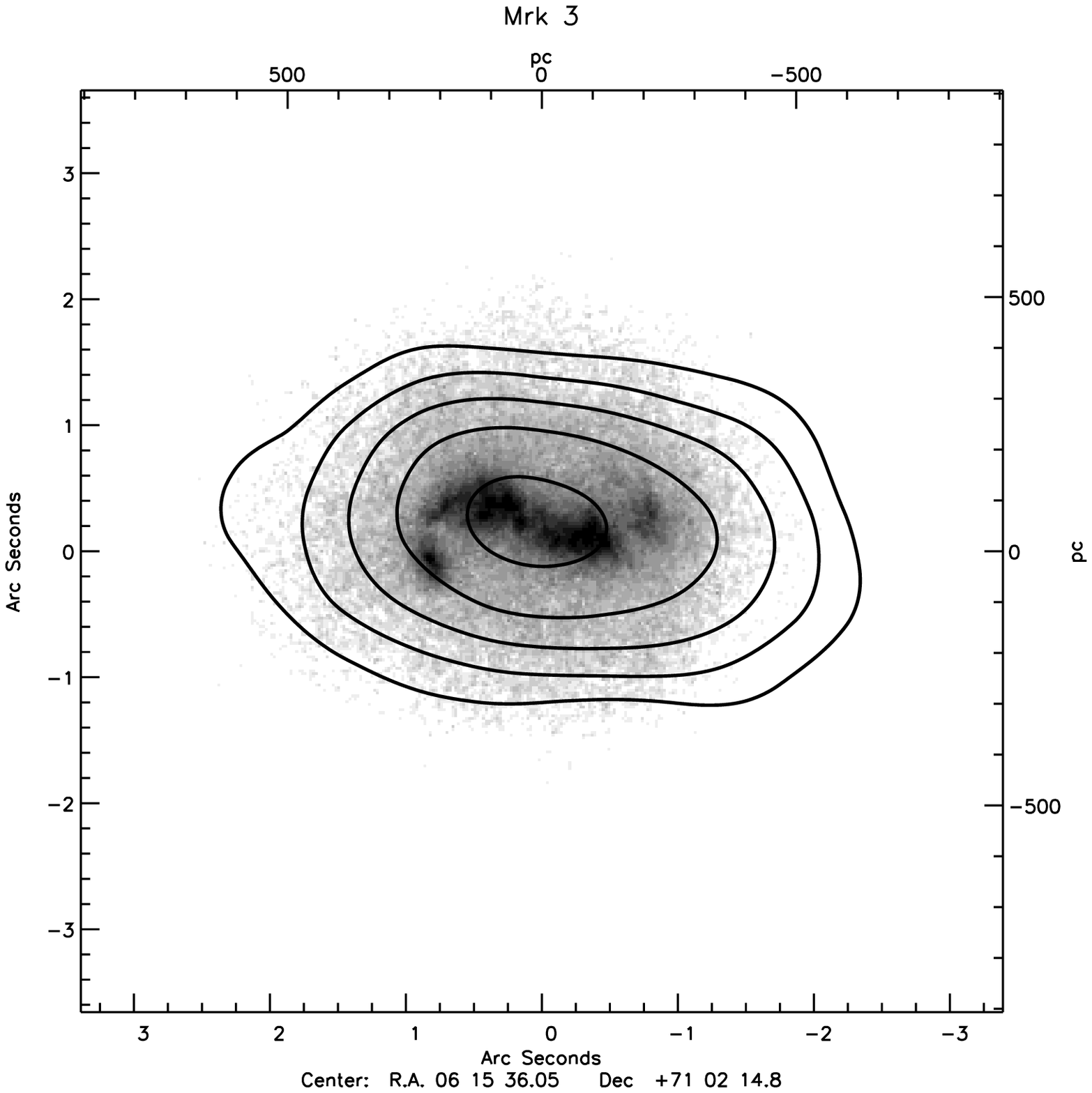, width=6.92cm}

\vspace{0.5cm}

\epsfig{file=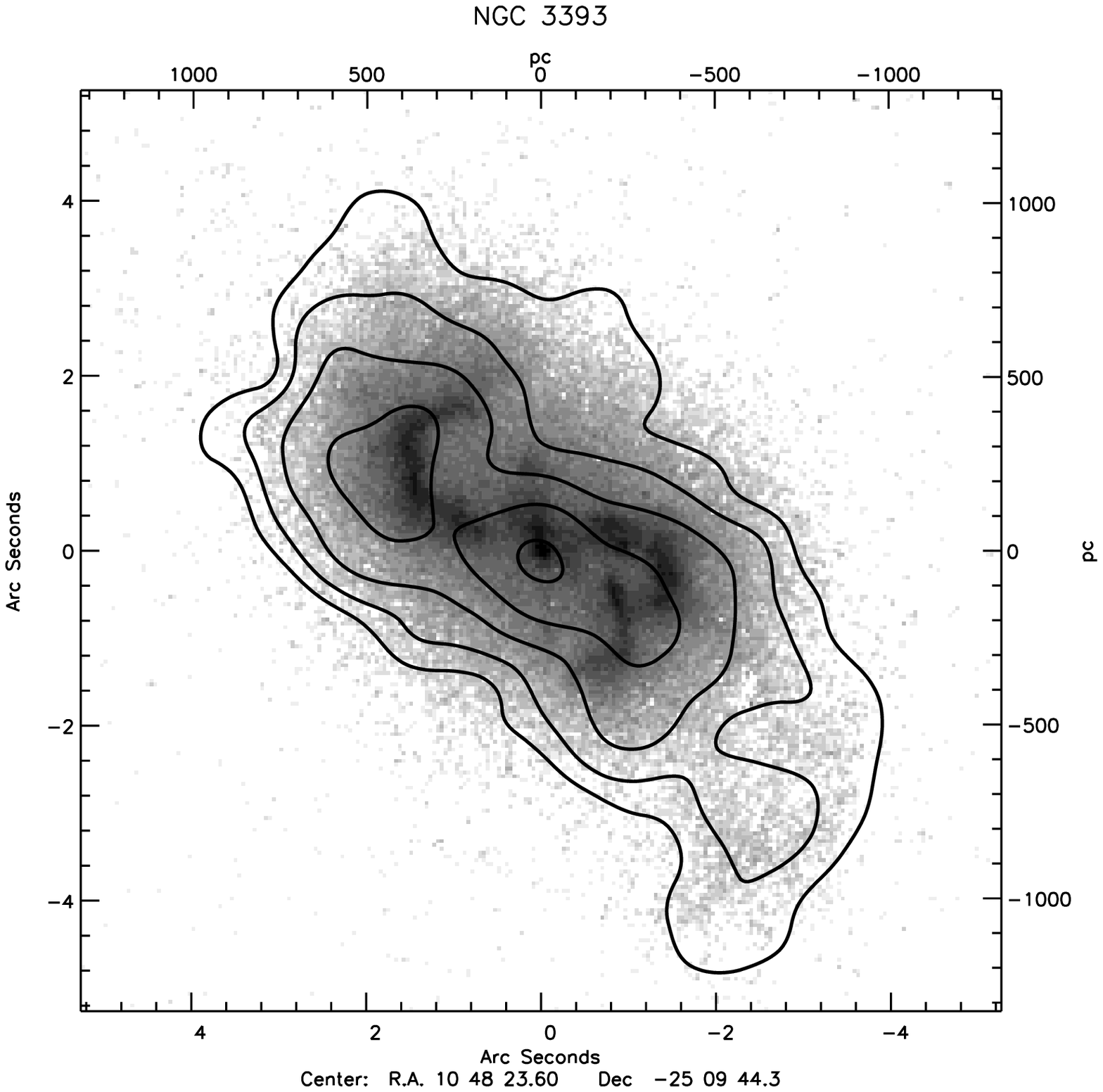, width=6.92cm}
\hspace{0.5cm}
\epsfig{file=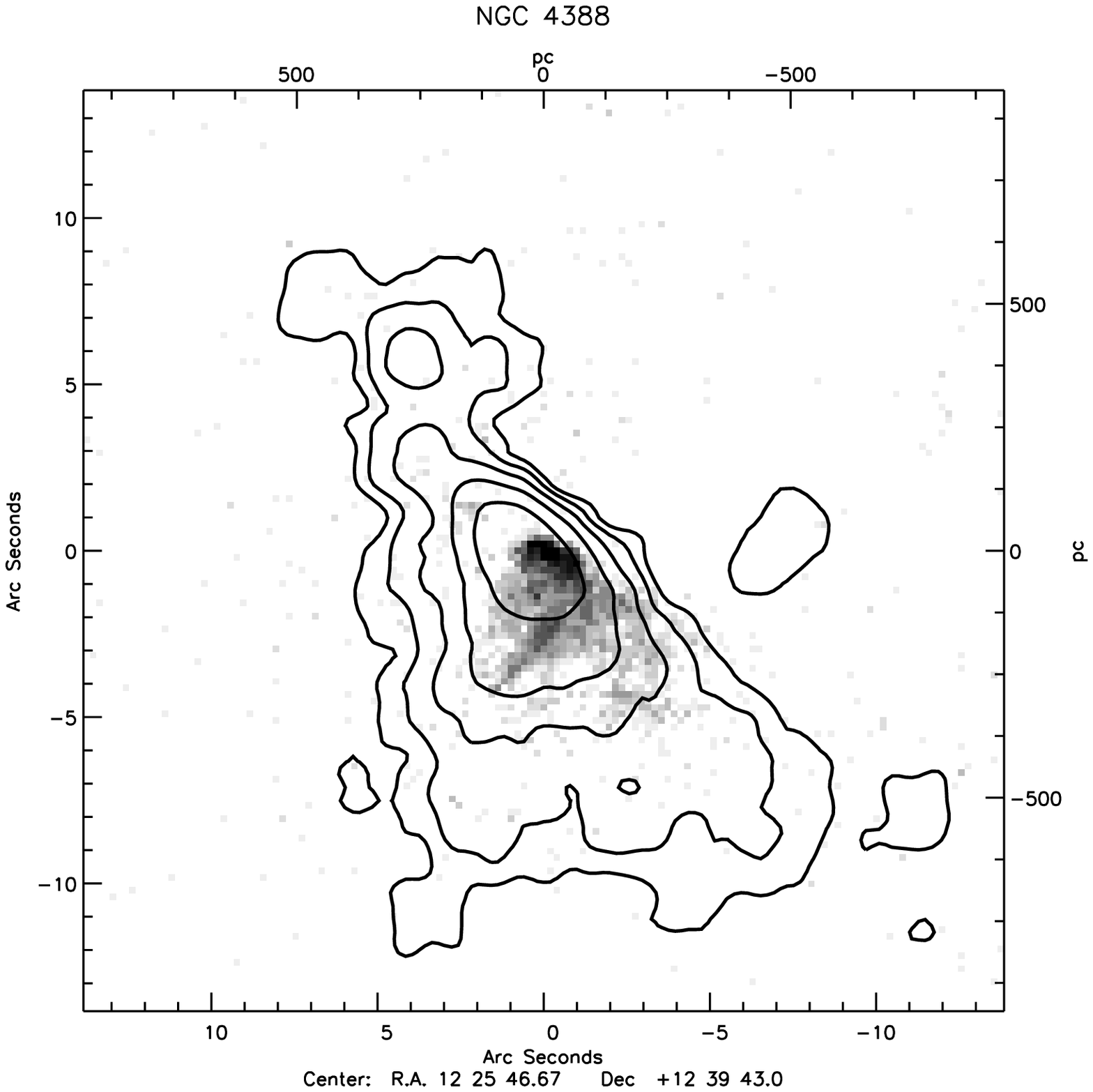, width=6.92cm}

\vspace{0.5cm}

\epsfig{file=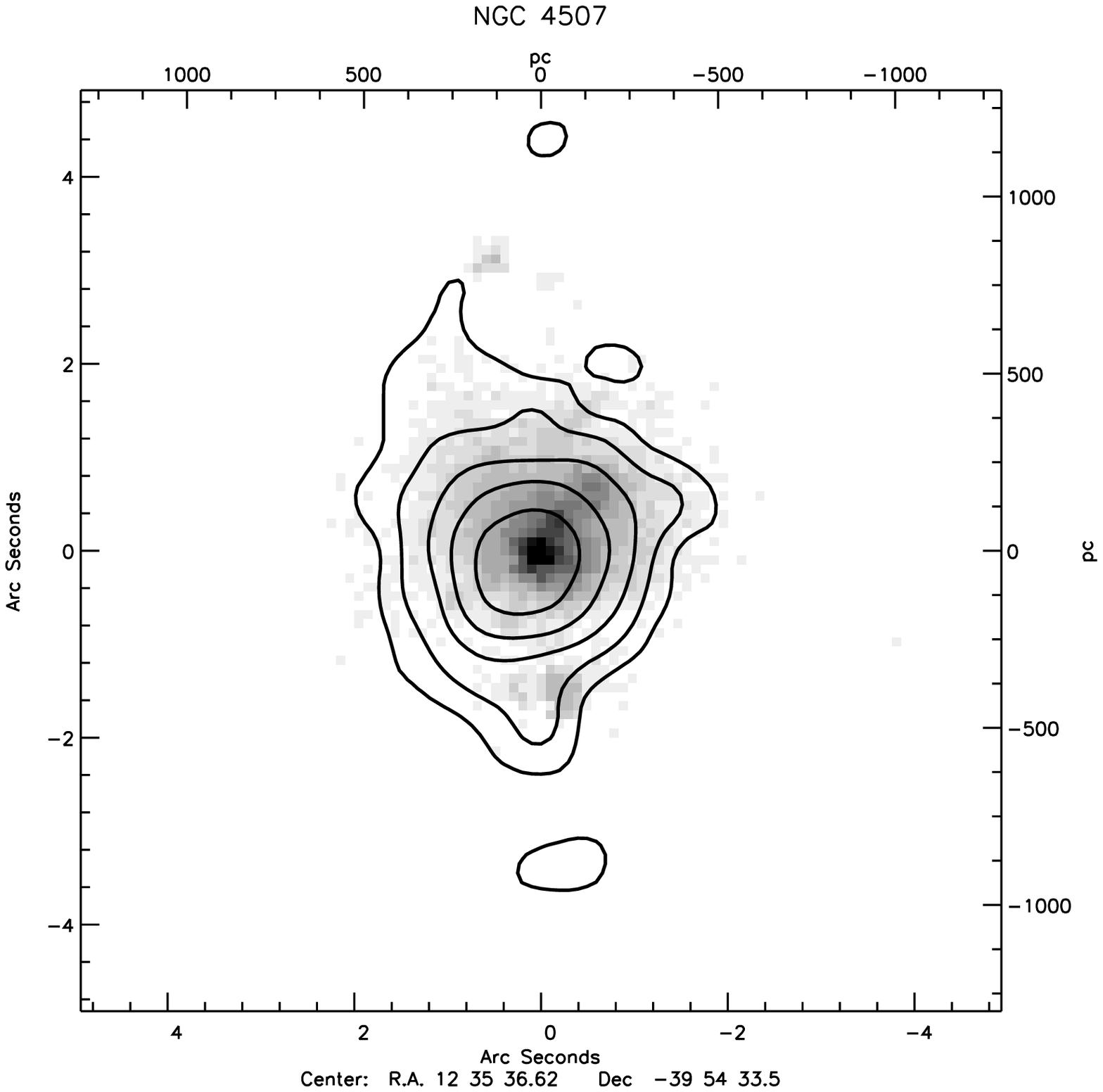, width=6.92cm}
\hspace{0.5cm}
\epsfig{file=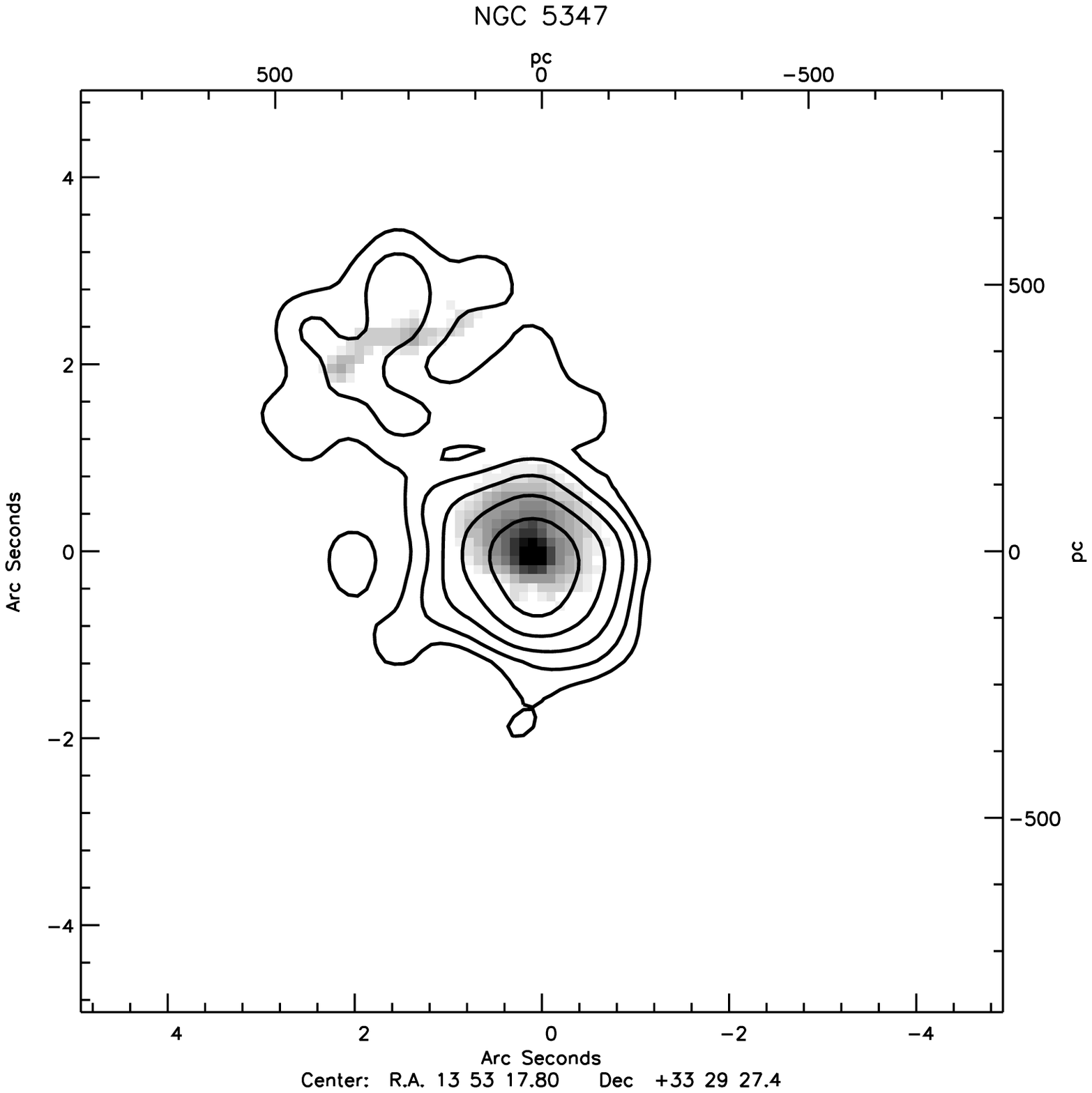, width=6.92cm}

\end{center}

\caption{\label{xray2oiiimaps}\textit{Chandra} soft X-ray contours superimposed on \textit{HST} [{O\,\textsc{iii}}] images. The contours correspond to five logarithmic intervals in the range of 1.5-50\% (NGC~1386), 5-80\% (Mrk~3), 5-90\% (NGC~3393), 4-50\% (NGC~4388), 0.5-50\% (NGC~4507) and 2-50\% (NGC~5347) of the peak flux. The \textit{HST} images are scaled with the same criterion for each source, with the exception of NGC~4388 and NGC~4507, whose [{O\,\textsc{iii}}] emission goes down to the 2\% and the 0.1\% of the peak, respectively. The images are aligned as explained in Sect. \ref{hst} and centered on the \textit{Chandra} count peak.}
\end{figure*}

Figure \ref{xray2oiiimaps} shows the contours of the \textit{Chandra} soft X-ray emission superimposed on the \textit{HST} [{O\,\textsc{iii}}] images, for all the sources in our sample. Since we are interested in the relationship between the X-ray and the [{O\,\textsc{iii}}] morphologies, we tried to adopt a similar dynamical range to display the two components, based on the peak flux of the image. The choice of the dynamical range to use for each source was basically dictated by the brightness of the nucleus in comparison to the weaker extended emission. This criterion, though being just one of the possible choices, allows us to have fairly homogeneous images for all the sample.

\begin{figure*}

\begin{center}

\epsfig{file=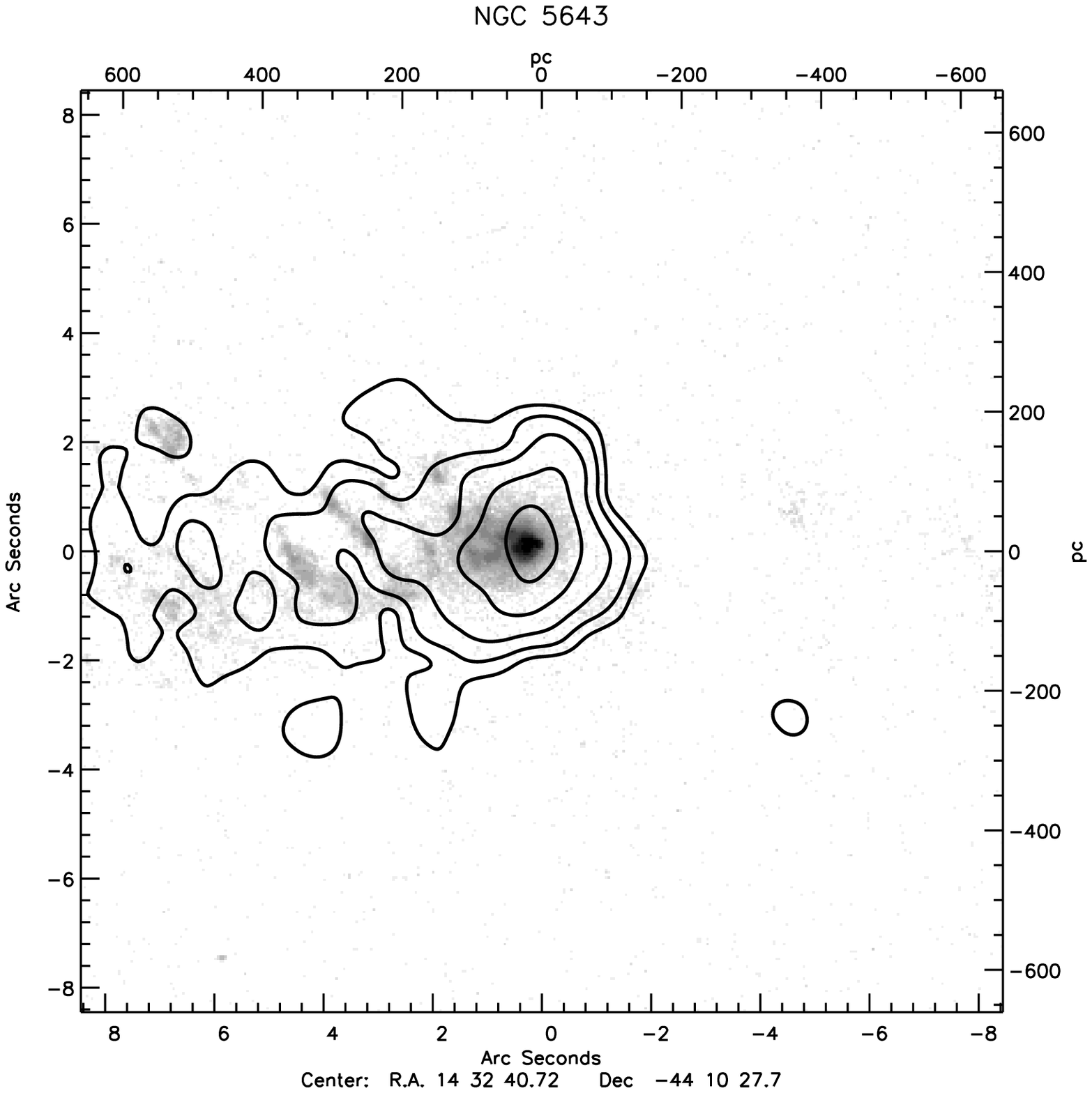, width=6.92cm}
\hspace{0.5cm}
\epsfig{file=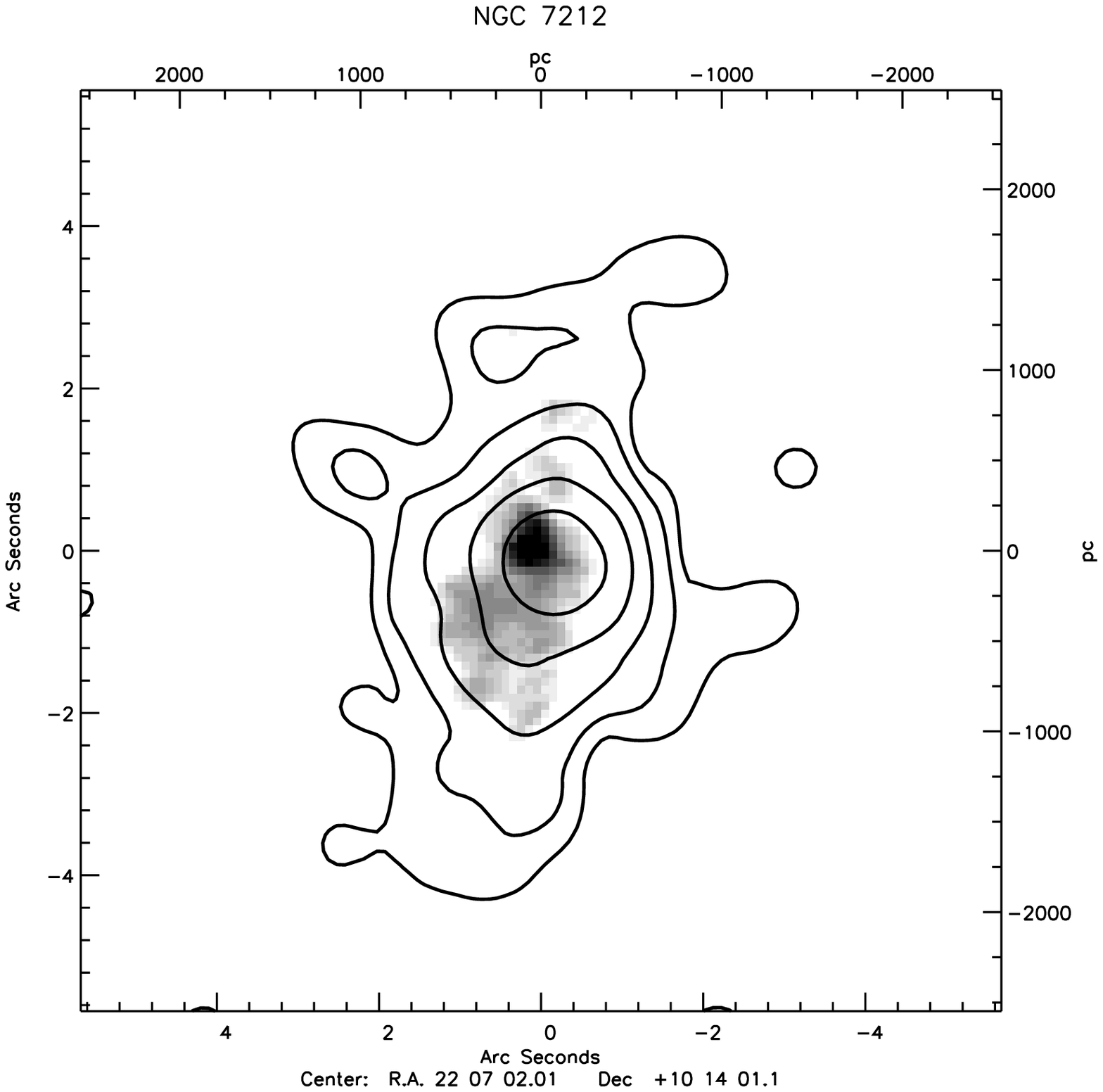, width=6.92cm}

\end{center}

\caption{\label{xray2oiiimaps_2}Same of Fig. \ref{xray2oiiimaps}, but for NGC~5643 (8-80\%) and NGC~7212 (1-50\%). In the case of NGC~5643, the [{O\,\textsc{iii}}] emission goes down to the 0.5\% of the peak.}
\end{figure*}

The coincidence between the soft X-ray and [{O\,\textsc{iii}}] emission is striking, both in the extension and in the overall morphology. Unfortunately, the lower spatial resolution of \textit{Chandra} with respect to \textit{HST} does not allow us to perform a detailed comparison of the substructures apparent in the latter. Moreover, the uncertainty in the relative alignment of the optical and X-ray images (see Sect. \ref{hst}) contribute to hinder this comparison, which would indeed be crucial to understand the real nature of the two components (see Sect. \ref{discussion}). Two sources deserve some further comments. In the case of NGC~4507, as already mentioned, the soft X-ray extension is not clear: however, the [{O\,\textsc{iii}}] emission is also extended on a very small scale and rather symmetrically. As for NGC~4388, it should be noted that the X-ray emission (which extends at least up to 16 kpc at the lowest energies) may well be influenced  by the Virgo cluster where the AGN lies. Moreover, this object may also be peculiar with respect to the others in our sample, because it presents also a Fe K$\alpha$ extended emission \citep{iwa03}.

Generally, the observed structures extend up to scales of kpc. While the morphologies for most sources could be approximately described as a bi-conical (or at least mono-conical) structure, for some objects a more complex representation is needed. In particular, Mrk~3 and NGC~3393 present S-shaped arms, which cannot be understood simply as collimation of the nuclear radiation by a torus. We will discuss the role of radio jets in the formation of these substructures in Sect. \ref{discussion}.

\subsection{\label{spectral}Spectral analysis}

As already mentioned in Sect. \ref{chandra}, we analyze in this paper for the first time the ACIS-S spectra of 5 out of the 8 objects in our sample. For the other three sources (namely Mrk~3, NGC~4388 and NGC~4507) we will refer to the results on their \textit{Chandra} spectra which have been already published in literature. However, in the case of Mrk~3, we will also present new results from a combined 190 ks XMM-\textit{Newton}/RGS spectrum.

In all the sources where the statistics were good enough to perform a comparison between the soft X-ray spectra from the nuclear and the extended region (as defined in Sect. \ref{chandra}), no significant differences were found between the two. In particular, the observed emission lines are required in both spectra. Therefore, all the following fits were performed on the total spectrum, which includes counts from both regions.

Two different models were tried to fit the soft spectra analyzed in this paper: a `thermal' model, reproducing emission from one or more optically thin, collisionally ionized plasma with free elemental abundances \citep[\textsc{Mekal} in \textsc{Xspec}:][]{mewe85}; a `scattering' model, i.e. a power-law and as many unresolved emission lines as required by the data, reproducing reflection of the nuclear continuum from a Compton-thin material photoionized by the AGN itself. In the latter model, the photon index of the powerlaw is left frozen to the value of $\Gamma=1.7$, which is a typical value for the primary continuum of Seyfert galaxies.

In the hard band, all the sources analysed in this paper are dominated by the iron emission line, while statistics are in most cases too poor to give significant constraints on the underlying continuum. However, when an additional component is required to model the hard X-ray spectrum, it will be explicitly reported in the text. As a consequence of the quality of the hard X-ray spectra of the sources, the iron line properties have been measured through the `local' fits (described in Sect. \ref{datared}): the values reported in Table \ref{fluxes} refer to these fits.

Detailed results on each source are described in the following subsections.

\begin{table}
\caption{\label{fluxes}Fluxes and iron line properties for the \textit{Chandra} observations analysed in this paper. The iron line fluxes and EWs refer to the `local' fits (see text for details).}

\begin{center}
\begin{tabular}{ccccc}
\hline
\textbf{Source} & \textbf{F}$_\mathrm{0.5-2\,keV}$ $^a$ & \textbf{F}$_\mathrm{2-10\,keV}$ $^a$ & \textbf{EW}$_\mathrm{Fe}$ $^b$ & \textbf{I}$_\mathrm{Fe}$ $^{c}$\\
\hline
&&&&\\
NGC~1386 & $1.8^{+1.5}_{-0.5}$ & $2.4\pm1.0$ & $1.5^{+0.8}_{-0.5}$ & $0.6^{+0.3}_{-0.2}$\\
 &  & & &\\
NGC~3393 & $2.2^{+0.5}_{-0.2}$ & $2.6\pm0.8$ & $0.6\pm0.3$ & $0.4\pm0.2$\\
 &  &  &  &\\
NGC~5347 & $0.27\pm0.06$ & $2.4\pm0.7$ & $1.7\pm0.7$ & $0.5\pm0.2$\\
 &  &  &  &\\
NGC~5643 & $1.4\pm0.2$ & $6.3^{+2.1}_{-2.0}$ & $1.5^{+0.8}_{-0.6}$ & $1.3^{+0.7}_{-0.5}$\\
 &  &  &  & \\
NGC~7212 & $0.8^{+0.1}_{-0.2}$ & $5.2^{+1.3}_{-1.2}$ & $0.7\pm0.3$ & $0.7\pm0.3$\\
 &  &  &  &\\
 \hline
\end{tabular}
\end{center}
$^a$In units of $10^{-13}$ erg cm$^{-2}$ s$^{-1}$

$^b$Iron line EW in units of keV

$^c$Iron line flux in units of $10^{-5}$ ph cm$^{-2}$ s$^{-1}$
\end{table}

\begin{figure*}

\begin{center}

\epsfig{file=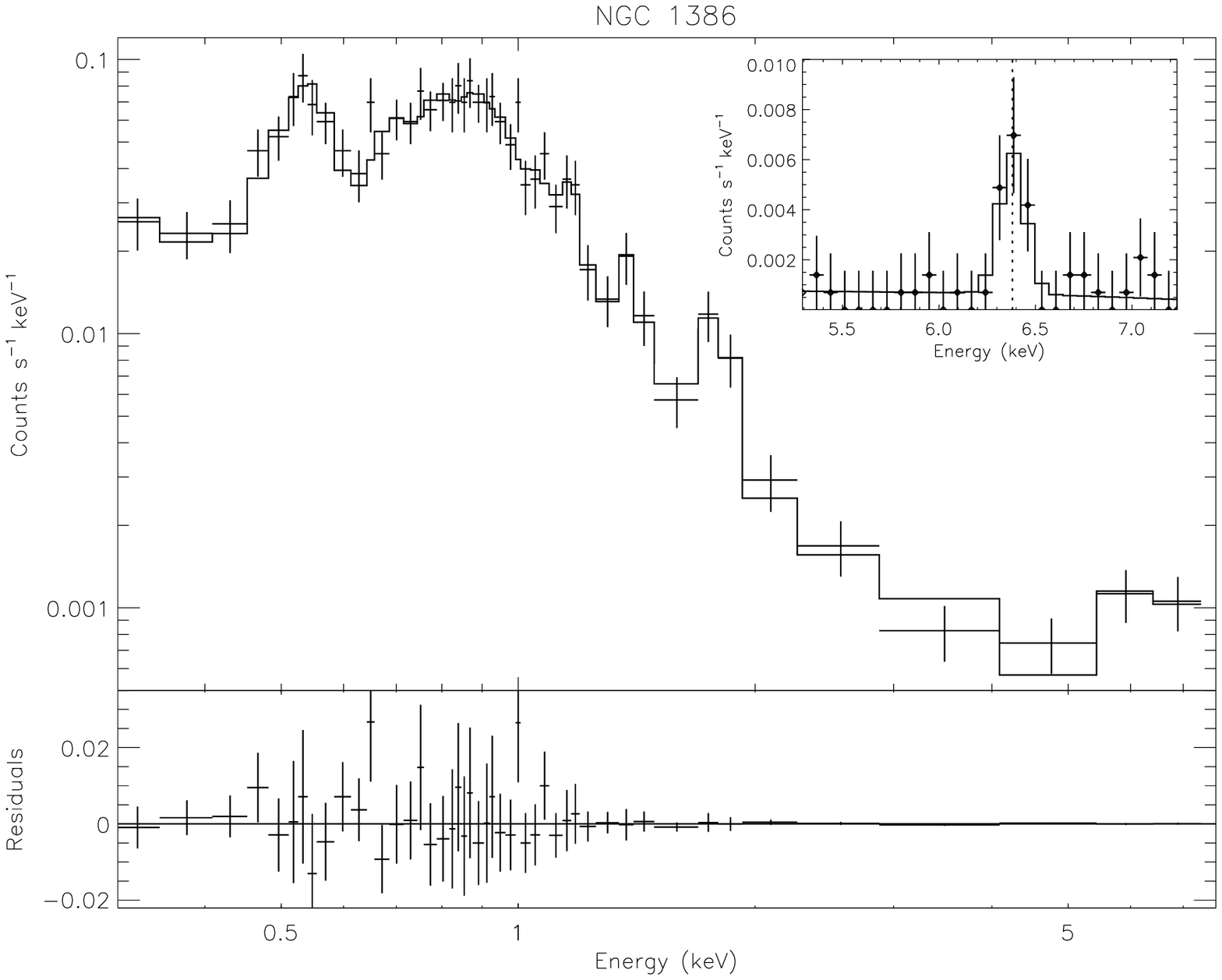, width=8cm}
\hspace{0.5cm}
\epsfig{file=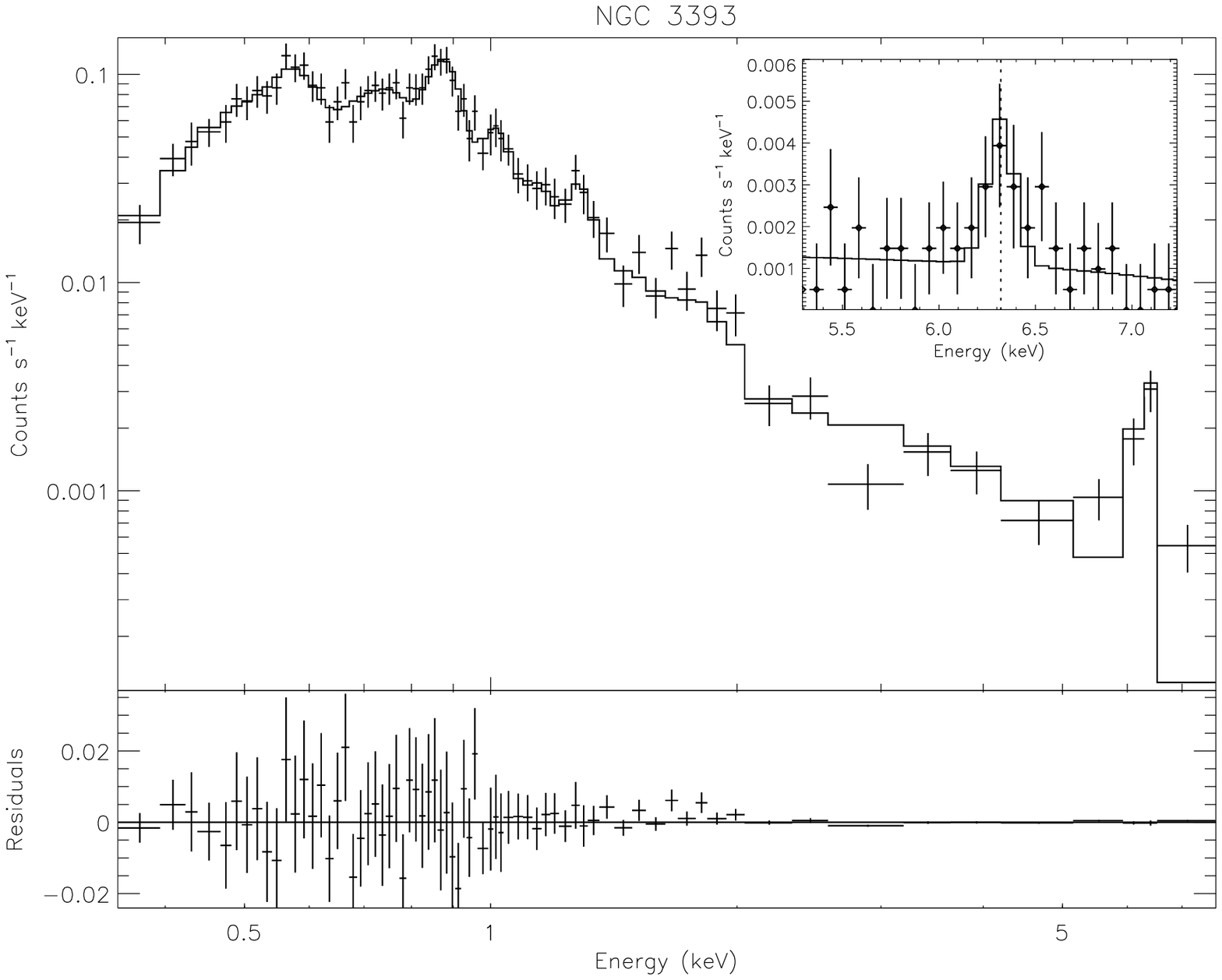, width=8cm}

\vspace{0.5cm}

\epsfig{file=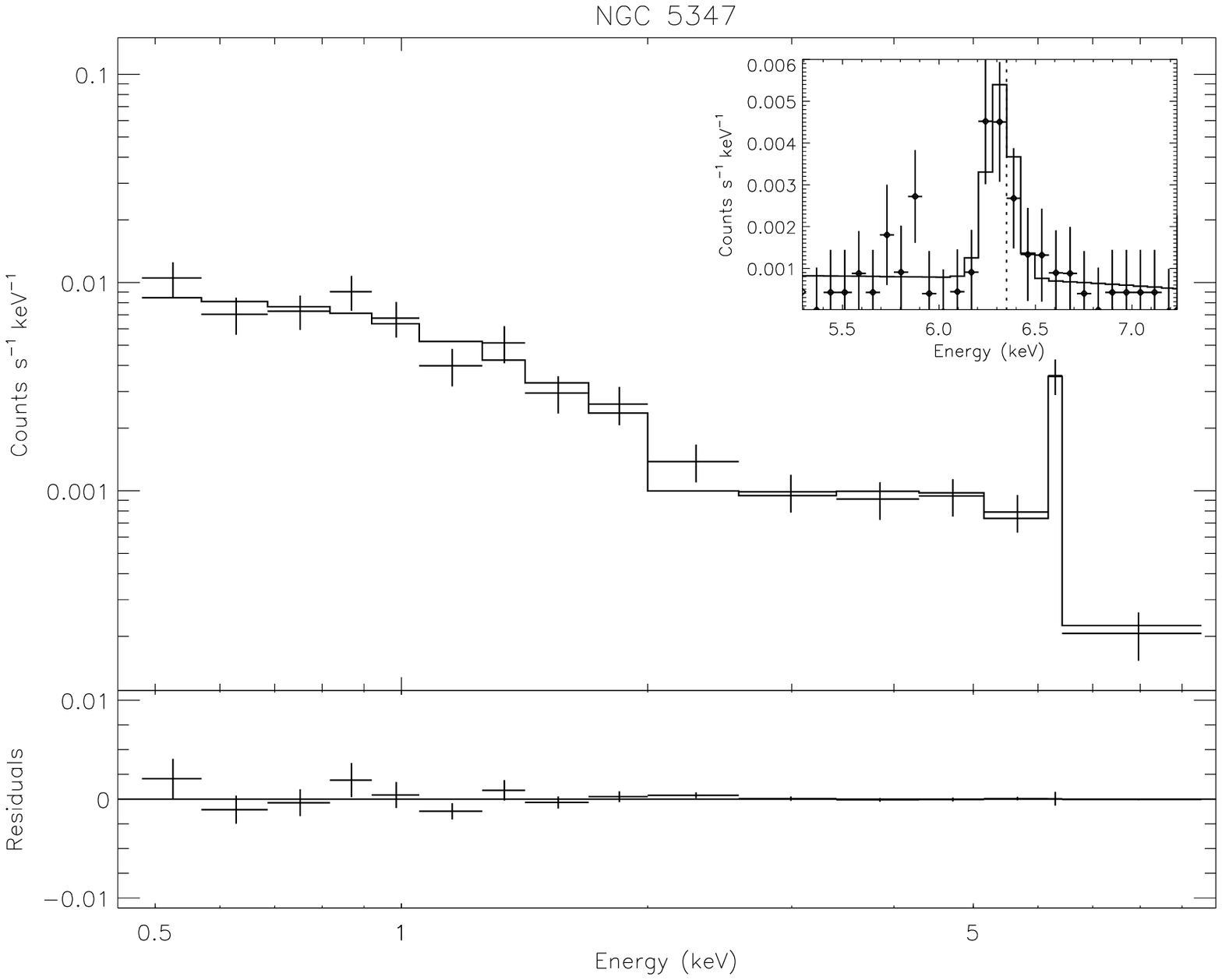, width=8cm}
\hspace{0.5cm}
\epsfig{file=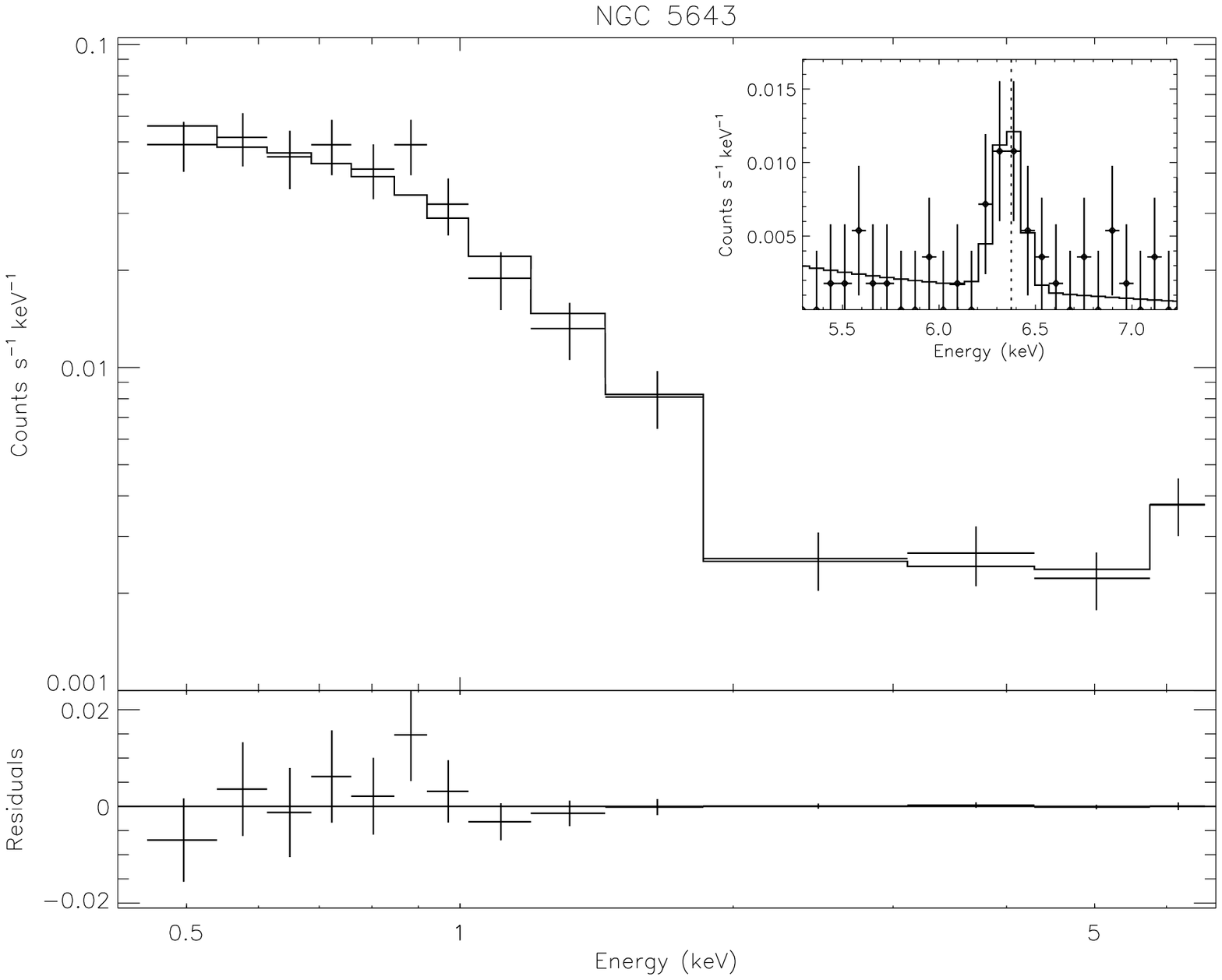, width=8cm}

\vspace{0.5cm}
\epsfig{file=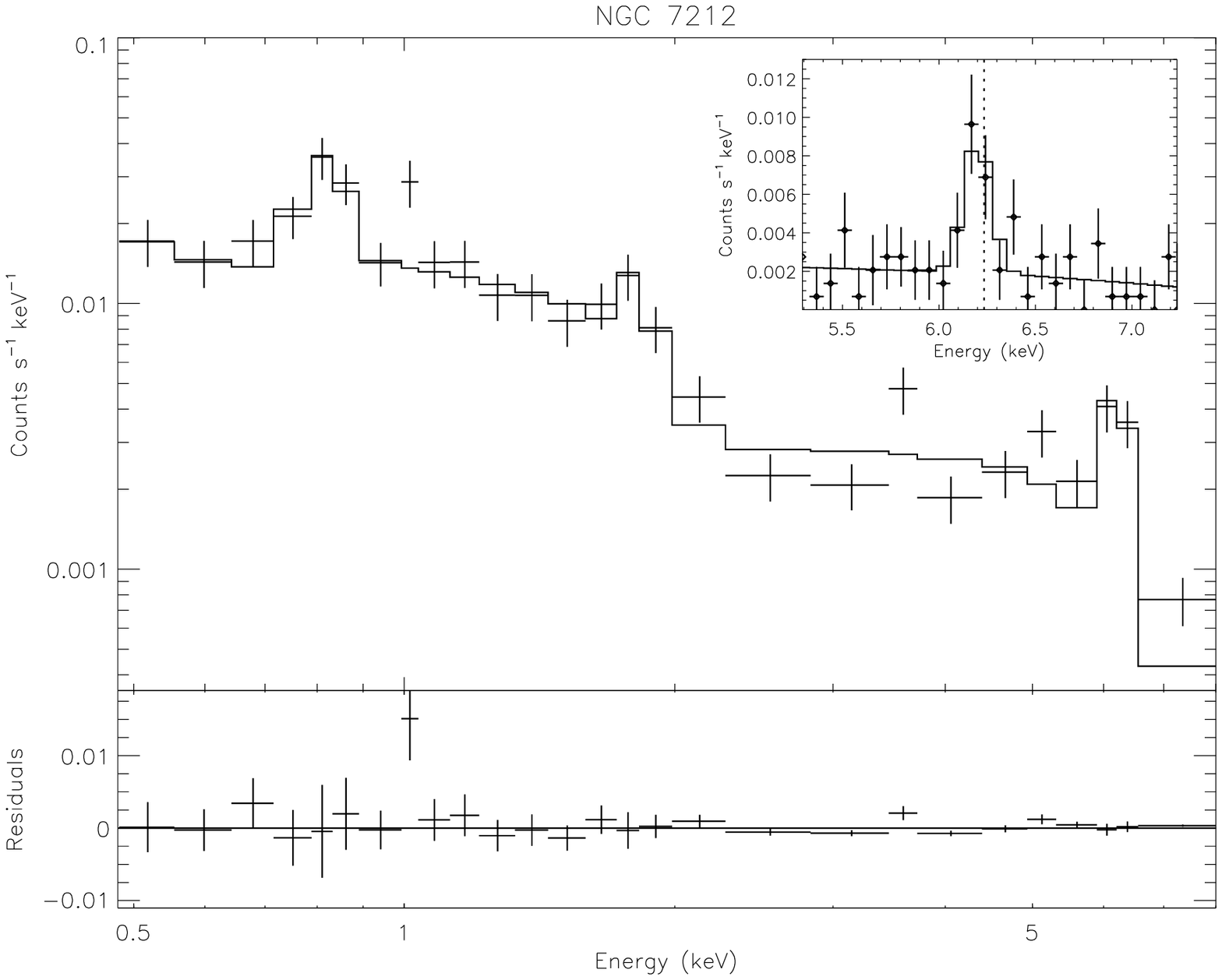, width=8cm}
\end{center}

\caption{\label{ironfits}\textit{Chandra} data and best fit models for the sources in our sample analysed for the first time in this paper. The inner box shows the `local' fit for the iron line, along with a dotted line to show the energy of the neutral iron K$\alpha$ line at the redshift of the source.}
\end{figure*}

\subsubsection{NGC~1386}

The source was found to be a reflection-dominated object in the XMM-\textit{Newton} observation \citep{gua05}. This is confirmed by the strong iron line in the \textit{Chandra} spectrum (see Fig. \ref{ironfits} and Table \ref{fluxes}), which by far dominates the hard spectrum and whose properties are fully consistent with those found with XMM-\textit{Newton}. Moreover, both the soft and the hard X-ray fluxes are in agreement with the ones reported by \citet{gua05}.

The EPIC soft X-ray spectra were best fitted with two thermal components \citep{gua05}. However, this model, though leading to values similar to those found with XMM-\textit{Newton} (temperatures of $0.14^{+0.05}_{-0.06}$ and $0.72^{+0.06}_{-0.07}$ keV, with a metal underabundance of $0.07^{+0.04}_{-0.03}$), fails to produce a good fit on the \textit{Chandra} data ($\chi^2=62/38$ d.o.f.). A much better fit is achieved with the `scattering' model ($\chi^2=18/20$). The lines included in the model are from H-- and H-like N, O, Ne, Mg, possibly blended with Radiative Recombination Continua (RRC) from C and O (see Table \ref{1386linespectrum}). The latter features, if confirmed by future, high resolution spectra of NGC~1386, would strongly favour an origin of this emission from photoionized plasma, like in the cases of NGC~1068 \citep{kin02,brink02} and Mrk~3 \citep{sako00b,bianchi05b}.

A final comment must be spent on a feature found at 0.29 keV. While it is possibly associated with the {C\,\textsc{v}} K$\alpha$ triplet, its large flux (of the order of $10^{-3}$ ph cm$^{-2}$ s$^{-1}$) makes its identification as an emission line troublesome. However, the feature lies at the end of the band considered `good' for ACIS spectra: below 0.3 keV calibration and background issues prevent us from performing a reliable analysis.

\begin{table}
\caption{\label{1386linespectrum}NGC~1386: emission lines included in the fit of the soft X-ray spectrum, along with their most likely identifications (see text for details).}

\begin{center}
\begin{tabular}{ccc}
\hline
\textbf{Energy} & \textbf{Flux} & \textbf{Id.} \\
(keV) & ($10^{-5}$ ph cm$^{-2}$ s$^{-1}$) &  \\
\hline
 &  &  \\
$0.53^{+0.01}_{-0.05}$ & $4.6^{+0.8}_{-1.5}$ & {N\,\textsc{vii}} K$\alpha$ - {O\,\textsc{vii}} K$\alpha$ \\
 &  &  \\
$0.66\pm0.02$ & $1.3\pm0.4$ & {O\,\textsc{viii}} K$\alpha$ - {O\,\textsc{vii}} K$\beta$\\
 &  &  \\
$0.76\pm0.01$ & $1.5\pm0.4$ & {O\,\textsc{vii}} RRC - {O\,\textsc{viii}} K$\beta$ \\
 &  &  \\
$0.84^{+0.02}_{-0.01}$ & $1.5\pm0.3$ & {O\,\textsc{viii}} RRC \\
 &  &  \\
$0.94^{+0.01}_{-0.02}$ & $1.1\pm0.3$ & {Ne\,\textsc{ix}} K$\alpha$\\
 &  &  \\
$1.05^\pm0.02$ & $0.6\pm0.2$ & {Ne\,\textsc{x}} K$\alpha$ \\
 &  &  \\
$1.16^{+0.02}_{-0.01}$ & $0.5\pm0.2$ & {Fe\,\textsc{xxiv}} L \\
 &  &  \\
$1.38\pm0.04$ & $0.2\pm0.1$ & {Mg\,\textsc{xi}} K$\alpha$\\
 &  &  \\
$1.78\pm0.05$ & $0.2\pm0.1$ & {Si\,\textsc{xiii}} K$\alpha$\\
 &  &  \\
\hline
\end{tabular}
\end{center}

\end{table}

\subsubsection{\label{mrk3}Mrk~3}

The \textit{Chandra} ACIS-HETG and the XMM-\textit{Newton} EPIC and RGS spectra were analyzed in detail by \citet{sako00b} and \citet{bianchi05b}, respectively. It is one of the few Compton-thick Seyfert galaxies that have high resolution spectra, showing that the soft X-ray spectrum is indeed dominated by emission lines. The ratio between the forbidden to the resonant line in the He-like triplets of light metals and the presence of RRC from the same elements give strong evidence in favour of an emitting gas photoionized by the nuclear continuum.

\begin{figure*}

\begin{center}
\epsfig{file=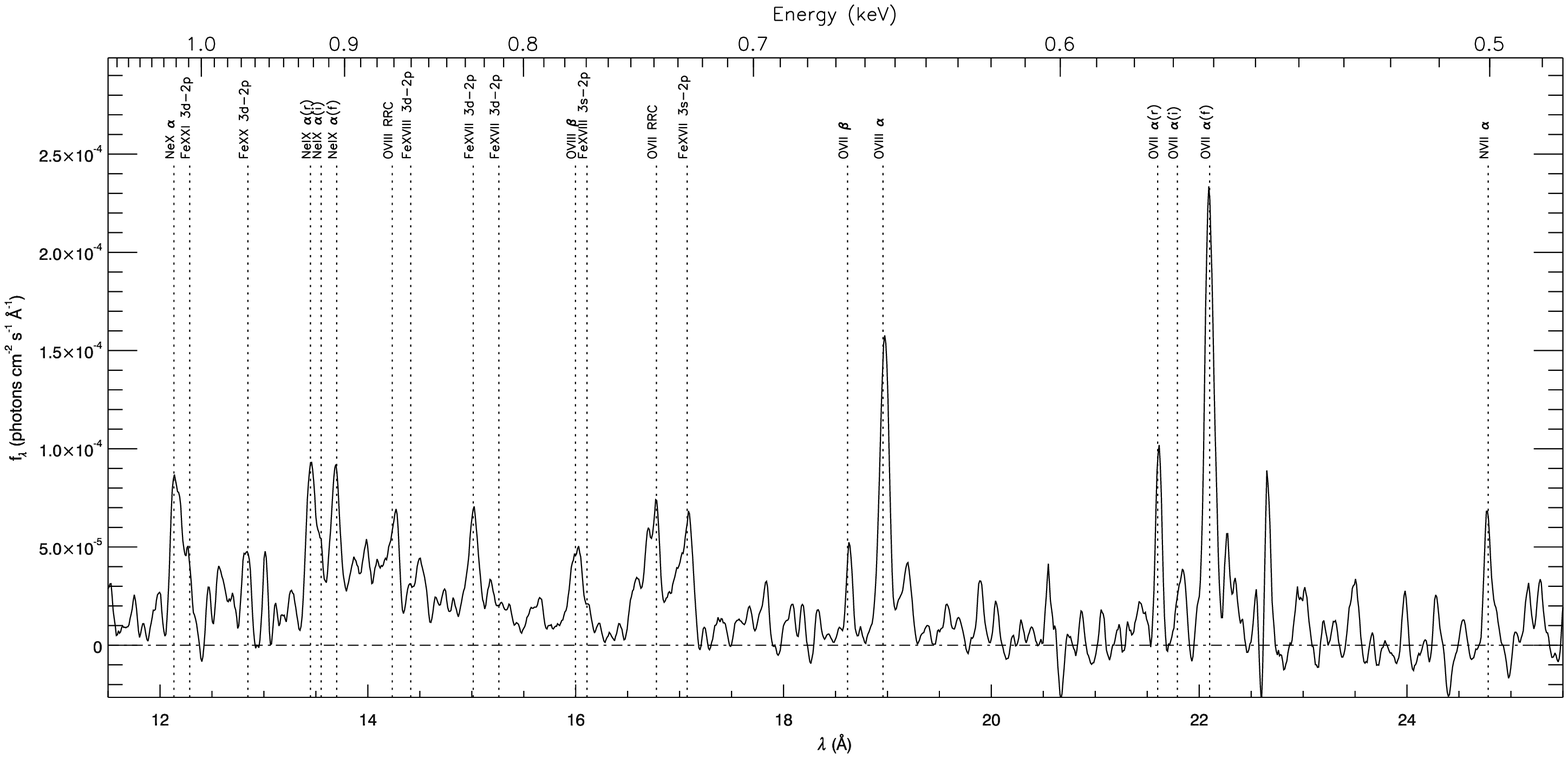, width=17.2cm}
\end{center}

\caption{\label{mrk3rgs}Mrk~3: XMM-\textit{Newton} 190 ks combined RGS spectrum, plotted in the rest frame of the source (12-25 $\AA$). The strongest emission features are labelled.}
\end{figure*}

We show in Fig. \ref{mrk3rgs} the high resolution spectrum with the largest S/N available for Mrk~3 and one of the best high resolution spectra for an AGN ever measured. It was obtained summing all the RGS observations of the source (for a total of $\simeq190$ ks) and combining RGS 1 and RGS 2 first and second orders.

The spectrum is dominated by K-shell emission lines from O and Ne, with weaker Fe L lines mainly from {Fe\,\textsc{xvii}} and {Fe\,\textsc{xviii}}. Spectra from plasma in collisional ionization equilibrium (CIE) are instead dominated by Fe L emission lines \citep[see e.g. the spectrum of solare flares: ][]{phil82}. As already observed by \citet{sako00b}, this could in principle be accounted for by an at least one order-of-magnitude iron underabundance, but this is at odds with the EW of the iron K$\alpha$ line and the depth of the neutral iron edge, which are consistent with an iron abundance only slightly lower than the solar one \citep{bianchi05b}. Moreover, the clear detection of narrow {O\,\textsc{vii}} and {O\,\textsc{viii}} RRCs implies an electron temperature much cooler than the one expected in a CIE plasma with the same ionic species \citep[e.g.][]{lp96,lied99}. Therefore, the gas is likely to be in photoionization equilibrium (PIE).

Indeed, the spectrum is telling us something more. The resonance lines in the {O\,\textsc{vii}} and {Ne\,\textsc{ix}} triplets are stronger than those expected for pure recombination in PIE \citep[e.g.][]{pd00}. The resonance 3\textit{d}-2\textit{p} Fe L lines should also be suppressed when recombination is the dominant process \citep{lied90}, while these transitions are clearly detected in the observed spectrum together with 3\textit{s}-2\textit{p} lines. Both problems can be solved taking into account the role of resonant scattering \citep{band90,matt94,kk95}. Resonant lines in He-like triplets are greatly enhanced by this mechanism with respect to recombination, especially for low column densities \citep[see e.g.][]{bianchi05}. Similarly, photoexcitation significantly favors the production of Fe L 3\textit{d}-2\textit{p} lines \citep{kall96}.

While a complete and detailed analysis of the combined RGS spectrum is beyond the scope of this paper, we compared the {O\,\textsc{vii}}-{O\,\textsc{viii}} series with the photoionization code \textsc{Photoion} \citep{kin03}, in order to get some more quantitative information from this high S/N part of the spectrum\footnote{As already reported by \citet{kin02}, \textsc{Photoion}, though being a local model in \textsc{Xspec}, does not allow a real fit procedure, because its complex nature makes it very difficult to deal properly with error calculations. Therefore, our `best fit values' have been found by trial and error, and we cannot estimate an error.}. We first only dealt with the spectrum between 0.55-0.66 keV (observer's frame), in order to include only the {O\,\textsc{vii}} triplet and the {O\,\textsc{viii}} Ly$\alpha$. The strong resonant line with respect to the forbidden and the intercombination lines is best reproduced with a {O\,\textsc{vii}} column density of  $7\times10^{17}$ cm$^{-2}$ and a transverse velocity distribution $\sigma_{v}^\mathrm{rad}=250$ km s$^{-1}$. Indeed, both parameters contribute to make resonant scattering an important process in the gas under analysis \citep[e.g.][]{kin02,bianchi05}. The observed forbidden line appears stronger than expected with respect to the intercombination line: this was also observed in the spectrum of NGC~1068 \citep{kin02, brink02} and explained with an additional contribution due to inner-shell photoionization in  {O\,\textsc{vi}} \citep[see also][for further details]{kin03}. The strong {O\,\textsc{viii}} Ly$\alpha$ line requires a much higher column density for this ion ($1.6\times10^{18}$ cm$^{-2}$), thus suggesting that the average ionization parameter could be high.

The inclusion of the spectrum up to 0.79 keV (observer's frame) shows that the flux of lower order lines ({O\,\textsc{vii}} and {O\,\textsc{viii}} $\beta$) and the {O\,\textsc{vii}} RRC is underpredicted by the parameters used to account for the $\alpha$ lines. Again, this is another piece of evidence in favour of a strong contribution from photoexcitation and, in general, an important indicator of photoionization as the dominant process at work \citep[see][for a detailed discussion on this issue]{kin02}. In order to reproduce correctly also these lines, we need lower column densities ($3\times10^{17}$ and $8\times10^{17}$ cm$^{-2}$ for {O\,\textsc{vii}} and {O\,\textsc{viii}}, respectively) and a lower $\sigma_{v}^\mathrm{rad}=200$ km s$^{-1}$. Note that it is not easy to disentangle the role of each of the two parameters separately, since both play on the relative role between recombination and photoexcitation. With these parameters, the need for an additional contribution for the {O\,\textsc{vii}} forbidden line flux is higher. Moreover, the observed {O\,\textsc{vii}} RRC keeps on being stronger than the expected one. We do not have an easy explanation for this behaviour, but it is likely that the close {Fe\,\textsc{xvii}} 3\textit{s}-2\textit{p} line (which is \textit{not} modelled by \textsc{Photoion}) may contaminate the flux of the RRC.

In summary, we confirm previous results by \citet{sako00b} and \citet{bianchi05b}, but on stronger evidence: the soft X-ray spectrum of Mrk~3 is produced in a gas in photoionization equilibrium, with a significant contribution from resonant scattering as well as recombination.

Moreover, visual inspection of Fig. \ref{mrk3rgs} also suggests that the gas is not strongly in/outflowing, since large shifts of the emission lines are not observed with respect to the theoretical values. This is quantitatively supported by the results on the spectral fits, which lead to tight constraints on the shifts of the {O\,\textsc{vii}} forbidden line, the {O\,\textsc{viii}} K$\alpha$ and the {N\,\textsc{vii}} K$\alpha$, being respectively $0\pm60$, $30\pm110$ and $10^{+180}_{-110}$ km s$^{-1}$ (note that the systematic uncertainty of the RGS at these energies is around 100-130 km s$^{-1}$). However, this may be simply due to the fact that we are observing a spectrum which includes both sides of the bicone, thus naturally producing a zero net shift. In any case, the line widths are also all upper limits, being $<300$, $<400$ and $<1\,000$ km s$^{-1}$ for the above-mentioned lines, respectively.

\subsubsection{NGC~3393}

On the basis of its BeppoSAX and XMM-\textit{Newton} observations, \citet{gua05} concluded that NGC~3393 is a Compton-thick source, even if the iron line was barely detectable because of low statistics. Our local fit on the \textit{Chandra} spectrum confirms the presence of a strong iron line as the dominant component of the hard spectrum (with a flux consistent with the one observed in the EPIC spectra), in agreement with the reprocessing-dominated nature of the object.

We first tried to fit the soft X-ray spectrum with a thermal model, as suggested by the XMM-\textit{Newton} analysis \citep{gua05}. We get parameters very similar to those found in the EPIC spectra (two \textsc{Mekal} components, with temperatures of $0.17^{+0.01}_{-0.02}$ and $0.70^{+0.06}_{-0.03}$ keV and an underabundance of $0.06^{+0.02}_{-0.01}$ with respect to the solar one), but the fit is unacceptable ($\chi^2=103/58$ d.o.f.), mainly because of large residuals at the {O\,\textsc{vii}} and {Ne\,\textsc{ix}} K$\alpha$ energies. We get a much better fit with a photoionized model ($\chi^2=33/44$). The required lines are similar to the ones found for NGC~1386 (see Table \ref{3393linespectrum}). The presence of a line at $\simeq1.8$ keV, detected by \citet{gua05} in the XMM-\textit{Newton} spectrum and expected in our model as emission from {Si\,\textsc{xiii}} K$\alpha$, is not required by the \textit{Chandra} data, being a detection lower than 2 $\sigma$.

\begin{table}
\caption{\label{3393linespectrum}NGC~3393: emission lines included in the fit of the soft X-ray spectrum, along with their most likely identifications (see text for details).}

\begin{center}
\begin{tabular}{ccc}
\hline
\textbf{Energy} & \textbf{Flux} & \textbf{Id.} \\
(keV) & ($10^{-5}$ ph cm$^{-2}$ s$^{-1}$) &  \\
\hline
 &  &  \\
$0.41^{+0.04}_{-0.06}$ & $2.8^{+2.7}_{-2.3}$ & {C\,\textsc{v}} RRC - {N\,\textsc{vi}} K$\alpha$ \\
 &  &  \\
$0.48^{+0.02}_{-0.01}$ & $4.6\pm1.2$ & {N\,\textsc{vii}} K$\alpha$\\
 &  &  \\
$0.573\pm0.001$ & $9.2^{+1.7}_{-1.0}$ & {O\,\textsc{vii}} K$\alpha$\\
 &  &  \\
$0.67^{+0.02}_{-0.01}$ & $2.2^{+0.6}_{-0.5}$ & {O\,\textsc{viii}} K$\alpha$ - {O\,\textsc{vii}} K$\beta$\\
 &  &  \\
$0.77^{+0.01}_{-0.02}$ & $2.5^{+0.5}_{-0.4}$ & {O\,\textsc{vii}} RRC - {O\,\textsc{viii}} K$\beta$ \\
 &  &  \\
$0.89^{+0.01}_{-0.02}$ & $3.0\pm0.4$ & {O\,\textsc{viii}} RRC - {Ne\,\textsc{ix}} K$\alpha$ \\
 &  &  \\
$1.03^{+0.01}_{-0.02}$ & $0.9^{+0.1}_{-0.3}$ & {Ne\,\textsc{x}} K$\alpha$ \\
 &  &  \\
$1.16^{+0.02}_{-0.03}$ & $0.3^{+0.1}_{-0.2}$ & {Fe\,\textsc{xxiv}} L \\
 &  &  \\
$1.30^{+0.02}_{-0.03}$ & $0.3\pm0.1$ & {Mg\,\textsc{xi}} K$\alpha$\\
 &  &  \\
\hline
\end{tabular}
\end{center}

\end{table}

\subsubsection{NGC~4388}

\citet{iwa03} analyzed in detail both the image and the spectrum of the \textit{Chandra} observation of this source. They found a strongly absorbed ($N_H\simeq4\times10^{23}$ cm$^{-2}$) spectrum, together with an extended soft X-ray emission, whose most likely origin is in a gas photoionized by the central AGN. Interestingly, \citet{iwa03} suggested a common origin for the [{O\,\textsc{iii}}] and the soft X-ray emission. Moreover, they also reported that no significant variations are found in the spectral shape along the radius of the extended emission, so that the ionization parameter remains fairly constant, thus implying a density which decreases with the distance roughly like $r^{-2}$.

\subsubsection{NGC~4507}

The \textit{Chandra} spectrum of NGC~4507 was analyzed in detail by \citet{matt04b}, together with the XMM-\textit{Newton} data. The source is seen through a large column density ($N_H\simeq4\times10^{23}$ cm$^{-2}$), while the presence of a Compton-thick reflector is inferred by the presence of a Compton reflection component, a large iron K$\alpha$ line and its Compton shoulder. The soft X-ray emission is dominated by emission lines, likely produced in a photoionized material. Interestingly, \citet{matt04b} found also an absorption line from {Fe\,\textsc{xxv}}, thus suggesting also the presence of an highly ionized gas, as observed in a large number of Seyfert galaxies \citep[see e.g. ][and references therein]{bianchi05}.

\subsubsection{NGC~5347}

The \textit{Chandra} spectrum of this object is clearly reflection-dominated: a good fit ($\chi^2$=9/12 d.o.f.) is achieved with a Compton reflection component, a steep ($\Gamma\simeq2.4$) powerlaw and a neutral iron line with a large EW (see Table \ref{fluxes}). The low statistics do not allow us to analyze in detail the soft X-ray spectrum, which can be also modelled by two thermal components, with loosely constrained temperatures around 0.16 and 2 keV and quasi-zero abundances ($\chi^2$=8/9 d.o.f.).

\subsubsection{NGC~5643}

The XMM-\textit{Newton} observation suggested a complicated picture for NGC~5643, with a heavily absorbing column density which likely obscures part of its own Compton reflection at high energies and a soft X-ray spectrum dominated by mainly unresolved emission lines produced in a photoionized plasma. Other possibilities cannot be, however, ruled out \citep{gua04}. The hard X-ray flux measured with \textit{Chandra} is consistent with the one found with XMM-\textit{Newton}, while the soft fluxes, though marginally, are not in agreement.

The quality of the \textit{Chandra} spectrum of NGC~5643 is not good enough to perform a detailed analysis in the soft band. However, a reasonable broad band fit ($\chi^2$=5/10 d.o.f.) is achieved with a pure Compton reflection component, an iron line and a steep ($\Gamma>3$) powerlaw to model the softer part of the spectrum. The strong iron line is confirmed by the local fit (see Table \ref{fluxes}). In this source, the statistics do not allow us to check if this component may be reproduced by the blending of the emission lines expected from a photoionized plasma, even if the XMM-\textit{Newton} analysis revealed that this is indeed possible \citep{gua04}. However, the thermal model with two \textsc{Mekal} components gives a fit of similar quality ($\chi^2$=3/7 d.o.f.), with temperatures of 0.15 and 0.62 keV and a metal abundance lower than 0.08 times the solar.

\subsubsection{NGC~7212}

The soft spectrum is well fitted by the scattering model with two emission lines (at $0.86^{+0.08}_{-0.05}$ and $1.84^{+0.06}_{-0.08}$ keV) required by the data at the $93\%$ and $95\%$ confidence level according to F-test, likely identified as K$\alpha$ emission from {Ne\,\textsc{ix}} and {Si\,\textsc{xiii}}. Indeed, the bad modelling of these residuals is the main reason why a thermal model fails to provide a good fit ($\chi^2=18/13$ d.o.f.). The broadband fit includes also a pure Compton reflection component and a strong iron line, confirmed by the local fit (see Table \ref{fluxes}), thus suggesting that the source is Compton-thick. This is in agreement with the XMM-\textit{Newton} observation \citep{gua05b}, whose soft and hard X-ray fluxes are also consistent with those measured by \textit{Chandra}.

\section{\label{cloudy}Photoionization models}

The spectral analysis of the sources in our sample suggests that their soft X-ray spectra are likely dominated by emission lines produced in a material photoionized by the central AGN. On the other hand, the striking resemblance of [{O\,\textsc{iii}}] structures with the soft X-ray emission favors a common origin for both components. Therefore, since the NLR is generally believed to be also produced mainly by photoionization, we generated a number of models in order to investigate if a solution in terms of a single photoionized material to produce the optical NLR and the soft X-ray emission is tenable.

Calculations were performed with version 96.01 of \textsc{Cloudy}, last described by \citet{cloudy}. The adopted model is represented by a conical geometry, which extends from 1 to 350 pc from the nucleus, with temperature set by photoionization equilibrium under a typical AGN continuum \citep{korista97}. We produced a detailed grid of models, as a function of the following parameters: ionization parameter \textit{U} of the illuminated face of the gas, defined as:

\begin{equation}
U=\frac {\int _{\nu _{R}}^{\infty }\frac{L_{\nu }}{h\nu} d\nu}
{4\pi r^{2}cn_{e}}
\end{equation}

\noindent where $c$ is the speed of light, $r$ the distance of the gas from the illuminating source, $n_{e}$ the electron density and $\nu_{R}$ the frequency corresponding to 1 Rydberg; filling factor of the gas \textit{f}; density of the gas $n_{e}$. We assume a power-law radial dependence of the density, $n_e(r_0)\, \left( \frac{r}{r_0} \right) ^{-\beta}$, with $\beta=0$ (constant density) and varying between 1 and 2.4. We limit our models to a minimum total column density of $10^{20}$ cm$^{-2}$. Note that the radial column densities under analysis are always consistent with a negligible edge opacity \citep[see e.g.][]{kin02}.

\begin{figure}
\begin{center}
\epsfig{file=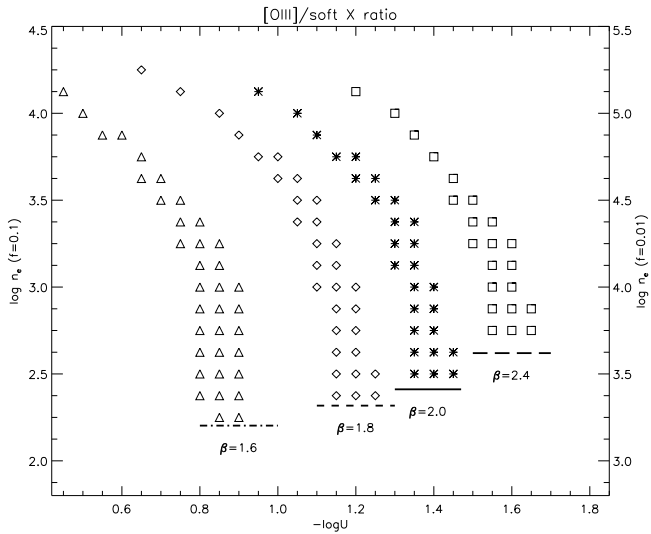, width=8.5cm}
\end{center}
\caption{\label{cone_ratio_r}Each symbol represents one solution in our grid of \textsc{Cloudy} models that satisfies the condition of total [{O\,\textsc{iii}}] to soft X-ray flux ratio in the range observed in our sample, plotted in a three-parameter space $U$, $n_e$ and $\beta$, i.e. the ionization parameter and the density at the beginning of the cone of gas (1 pc), and the index of the density powerlaw, represented by different symbols (\textit{triangles}: $\beta=1.6$, \textit{diamonds}: $\beta=1.8$, \textit{stars}: $\beta=2.0$, \textit{squares}: $\beta=2.4$). The horizontal lines determine the limit corresponding to a total column density of $10^{20}$ cm$^{-2}$ for each index: solutions below this limit are not plotted. Solutions with $\beta=2.2$ are not plotted for clarity reasons. The net effect of the filling factor $f$ is to shift the density values: the two y axes refer to $f=0.1$ and $f=0.01$ (see text for details).}
\end{figure}

The ratio between the [{O\,\textsc{iii}}] $\lambda5007$ line and the soft X-ray emission (defined as the total flux of the K$\alpha$ and K$\beta$ emission lines from N, O, Mg, S, Si in the range 0.5-2.0 keV) was calculated for each set of parameters. In Fig. \ref{cone_ratio_r}, each symbol represents a solution in the $U$ versus $n_e$ plane, where this ratio has a value within 2.8 and 11, as observed in our sample (see Table \ref{log}). Different symbols correspond to different values of $\beta$. The net effect of changing the filling factor is simply a shift of the solutions along the density axis, by a factor equivalent to the variation in $f$, thus reproducing the same total column density for each set of three parameters consituting a `good' solution. The solutions occupy well-defined regions in the $n_e-U$ diagram, with those with lower $\beta$ being at larger values of ionization parameters.

\begin{figure}
\begin{center}
\epsfig{file=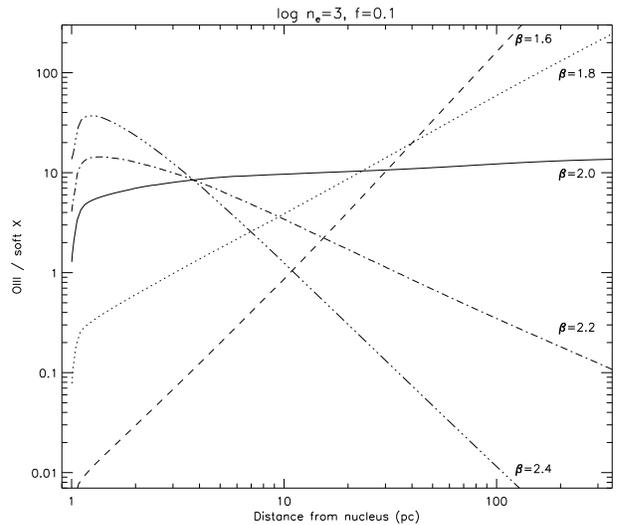, width=8cm}
\end{center}
\caption{\label{emiplot}The [{O\,\textsc{iii}}] to soft X-ray ratio plotted as a function of the radius of the gas, for different values of $\beta$, when $\log{n_e}=3$ and $f=0.1$. The corresponding values of $\log{U}$ for these solutions are (from top to bottom): -0.85, -1.15, -1.4, -1.5, -1.6.}
\end{figure}

The reason for this behaviour becomes clear inspecting Fig. \ref{emiplot}, where the [{O\,\textsc{iii}}] to soft X-ray flux ratio is plotted as a function of the radius of the gas. Since $U\propto n_e^{-1} r^{-2}$, all density laws with $\beta<2$ produce a gas with a ionization parameter decreasing along with the distance. In these cases ($\beta=1.6$ and 1.8 in Fig. \ref{emiplot}), most of the soft X-ray emission is produced in the inner radii of the cone, while the bulk of the [{O\,\textsc{iii}}] emission is produced farther away, where the gas is less ionized. If $\beta=2$, the ionization parameter remains fairly uniform up to large radii, so that the total observed ratio between [{O\,\textsc{iii}}] and soft X-ray is constant with radius. Finally, if $\beta>2$, the ionization parameter increases with radius, so most of the soft X-rays are actually produced at larger radii, while the [{O\,\textsc{iii}}] emission line is mostly concentrated around the nucleus (cases $\beta=2.2$ and 2.4 in Fig. \ref{emiplot}).

This is the reason why solutions with $\beta<1.6$ and $\beta>2.4$ are not plotted in Fig. \ref{cone_ratio_r}, even if these exist, satisfying the overall [{O\,\textsc{iii}}] to soft X-ray flux ratio. Their radial behaviour is radically different from what seen in Fig. \ref{xray2oiiimaps} and \ref{xray2oiiimaps_2}, where both the [{O\,\textsc{iii}}] and the soft X-ray emission are produced up to large radii. In particular, it is worth noting that constant-density solutions are totally unacceptable.

Before going through the discussion on the results of these calculations, we would like to point out some caveats. The flux included in the quantity we call `soft X-ray' does not include Fe L emission, which is known to be quite important, even if not dominant, in photoionized spectra (see, for example, Fig. \ref{mrk3rgs}). Moreover, no contribution from the scattered continuum has been taken into account in the total soft X-ray flux. However, this should not be very relevant \citep[see e.g.][]{kin02,bianchi05b}. In any case, the inclusion of these components is not expected to change dramatically the value of the [{O\,\textsc{iii}}] to soft X-ray ratio.

We have not investigated the effects of varying the incident continuum. In particular, some authors include X-ray and UV absorbers between the nucleus and the NLR clouds, finding evidence that an absorbed incident continuum is actually needed by the data \citep[see e.g.][]{alex99,krae00b}. The presence of this absorber would change the number of ionizing photons in the incident radiation on the NLR clouds, thus effectively changing the ionization equilibria calculated in our model. We will discuss the role and the properties of these absorbers in the next section, particularly regarding their possible link with the gas analysed in this work.

Finally, the role of dust was not included in our models. However, \citet{gal03} found that the mid-infrared (MIR) emission in NGC~1068 is very well correlated with the [{O\,\textsc{iii}}] NLR, thus suggesting the presence of dust grains in this material. The dust could have relevant effects on the absorption of optical emission, thus changing the value of the ratio between [{O\,\textsc{iii}}] and the soft X-ray flux.

As a consequence of these caveats, these models should be treated as simple toy models. In order for more complex models to be tested, one would need X-ray imaging detectors matching the HST sub-arcsec resolution and possibly space resolved high-resolution X-ray spectroscopy at the same scale, but both requirements will not be available for many decades to come. Our results are a proof of existence of simple scenarios consistent with the currently available experimental data. However, they do not imply that more complex or different scenarios can be ruled out, as we will discuss in detail in the next Section.

\section{\label{discussion}Discussion}

The analysis of the \textit{Chandra} images presented in this paper has unambiguously shown that there is a spatial correlation between the soft X-ray emission in Seyfert galaxies and their [{O\,\textsc{iii}}] emission. Moreover, the spectral analysis of the same sources suggests that the origin of the soft X-ray component is likely in a photoionized gas. In one case, Mrk~3 (but see also the cases of NGC~1068 and Circinus, not included in this sample), this component is actually resolved in plenty of emission lines in high resolution spectra and a number of diagnostic tools indicate that they are produced by gas in photoionization equilibrium. In all the other sources, where only CCD spectra are available, a thermal model is always disfavored, in most cases, on statistical grounds, but also because it leads to unphysically low metal abundances. In the latter case, however, a thermal component cannot be completely ruled out, since low abundances can be, in principle, mimicked by iron depletion from dust in interstellar clouds \citep{ptak97} or dilution by featureless continua, produced, for example, in binaries \citep{gua99}.

The photoionization models presented in the previous section suggest that it is in principle possible to have a gas, photoionized by an AGN and extending for hundreds of pc, which emits the right ratio of [{O\,\textsc{iii}}] to soft X-ray with a reasonable radial profile. These results imply peak densities around $10^3$ cm$^{-2}$ (for a filling factor of 0.1), which are reasonably in agreement with what found for the NLR of many sources \citep[see e.g.][]{cooke00,nw00,fsl03,bkb04}. Moreover, solutions with constant density must be rejected, and density laws that fall similarly to $r^{-2}$ have to be preferred.

\begin{figure}
\begin{center}
\epsfig{file=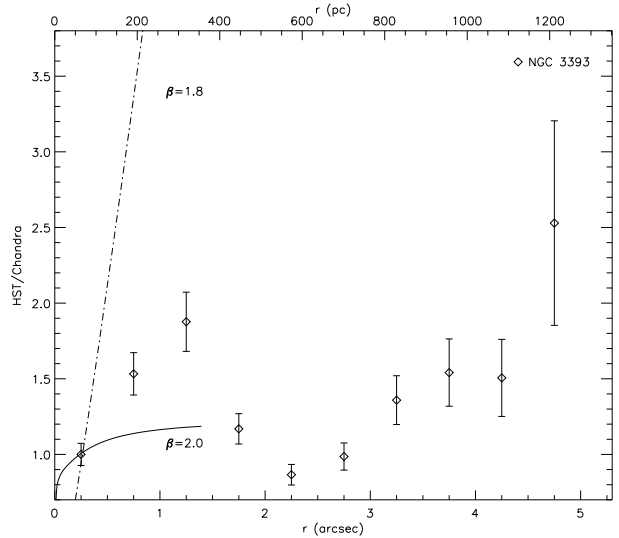, width=8cm}
\end{center}
\caption{\label{rprofile}NGC~3393: \textit{HST} to \textit{Chandra} soft X-ray counts ratio as a function of distance from the nucleus. The two curves refers to the photoionized models plotted in Fig. \ref{emiplot} (see text for details).}
\end{figure}

We plot in Fig. \ref{rprofile} the ratio between the \textit{HST} and the soft X-ray \textit{Chandra} counts as a function of the distance from the nucleus for the source with the best \textit{Chandra} data in our sample, NGC~3393. The counts at each radius are taken from rectangular regions 0.5 arcsec wide and 5 arcsec long, perpendicular to the axis of the extended region. The \textit{HST} image was convolved with a Gaussian with FWHM=0.8 arcsec to mimic \textit{Chandra} resolution. The plot shows that this ratio is fairly constant within a factor of a few, thus requiring an ionization parameter which changes slowly with radius. In particular, the data points in Fig. \ref{rprofile} suggest a decrease somewhat flatter than 2, but still steeper than $r^{-1.8}$. Interestingly, similar results are found for the NLR on the basis of the ratio of the [{Si\,\textsc{ii}}] $\lambda 6717$ to $\lambda 6731$, [{O\,\textsc{iii}}] $\lambda 5007$ to H$\beta$ and [{O\,\textsc{ii}}] $\lambda 3727$ to H$\beta$ for a number of sources, \citep[see e.g.][]{krae00b,kc00,bkb04}, including, notably, Mrk~3 \citep{coll05}.

A simple explanation for this density decrease with radius may be that the NLR clouds are not pressure confined and so they tend to expand outwards, as suggested by \citet{krae00b}. A density fall like $r^{-2}$ would arise naturally if the outflow had a constant velocity and mass rate. However, this picture is clearly oversimplified and other mechanisms can contribute to the observed radial behaviour of the ionization parameter. In particular, the [{O\,\textsc{iii}}] structures are generally found very well correlated with the radio images, thus suggesting an important role of jets in the overall NLR shape. We refer the reader to the papers where the radio images of these sources are presented to see the correlation with the [{O\,\textsc{iii}}]/soft X-rays components shown in Fig. \ref{xray2oiiimaps}: NGC~1386 \citep{nag99}, Mrk~3 \citep{cap99}, NGC~3393 \citep{cooke00}, NGC~4388 \citep{kuk95}, NGC~4507 \citep{morg99}, NGC~5643 \citep{mor85}, NGC~7212 \citep{fal98}. On the other hand, it is worth noting that NGC~5347 represents an exception, since it does not present extended radio emission \citep{thean00,schm03b}.

In particular, it has been proposed that the expansion of the radio ejecta compresses the gas, enhancing its radiative emission, thus effectively dominating the observed morphology of the NLR. Moreover, \citet{sbd93} have also suggested that the interaction between the radio plasma and the surrounding medium produce an ionizing continuum capable of photoionizing the same gas. In other words, the local ionization parameter in a particular region of the NLR may be affected by this interaction as well as from the AGN continuum. It is clear that this phenomenon would reduce the need for a steep decrease of the density with the radius, since local sources of photoionization would prevent the ionization parameter from falling steeply with radius.

We have already mentioned the presence of the X-ray absorber in the previous section. The so-called `warm absorbers' are often observed in the spectra of Seyfert 1s and are thought to be located between the Broad Line Region (BLR) and the torus \citep[see e.g.][and references therein]{blust05}. If this is the case, this material is not supposed to be an important emitter in Seyfert 2s or not even be observable because obscured by the torus. On the other hand, the gas under analysis in this paper may well be along the line of sight when observing a Seyfert 1 and so effectively add its column density to the inner warm absorber. The column densities of warm absorber are measured approximately in the range $10^{21}$ -- $10^{23}$ cm$^{-2}$, but are believed to be composed of various phases, possibly with different geometrical and kinematic properties \citep[e.g.][]{blust05}. In this sense, it is still possible that the denser and inner part of the soft X-ray emitting region resulting in the scenario discussed in this paper may represent the outer part of the warm absorber.

In this regard, it is interesting to compare the gas velocity as inferred from the RGS spectra of Mrk~3 with those typically found in the warm absorbers and in the NLR. As we have seen in Sect. \ref{mrk3}, the energy shifts measured in the X-ray emission lines of Mrk~3 imply that the gas has a very low bulk velocity, no larger than around 100 km s$^{-1}$. On the other hand, the warm absorbers are often found to be outflowing at velocities larger than a few hundreds km s$^{-1}$ \citep[e.g.][]{blust05}. This is, at least qualitatively, in agreement with the velocity maps measured for the NLR, which resemble those of the warm absorber in the region closer to the nucleus, then decreasing toward zero velocity outwards \citep[see e.g.][and references therein]{ruiz05}.

Indeed, it has been suggested that even the NLR could be made up of different components, with different density and ionization \citep[e.g.][]{kf89,krae00b}. We have clear evidence of the need of materials with different ionization stages also in X-rays, with the observation, in a large number of objects, of lines from highly ionized iron, inconsistent with the other lines from lighter metals observed in the soft X-ray spectra  \citep[see e.g.][]{bianchi05}. Moreover, all the solutions presented in this paper have negligible emission from Si, while H and He-like emission from this element are detected even in some of the sources included in our sample. In this sense, it is likely not surprising that the values of U of our solutions are larger than the ones typically required by the modelling of the NLR \citep[see e.g.][]{krae99,bl05}. The latter are probing the gas dominated by the optical emission, while ours are basically driven by the need of a strong X-ray component. Interestingly, a component with large U, consistent with what we find in our model, is also invoked in the NLR to produce optical high-ionization lines \citep{kf89,krae00b}.

An easy solution would be that all the components with different ionization parameters are spatially separated, possibly radially distributed, but the fact that the [{O\,\textsc{iii}}] to soft X-ray ratio presented in this paper remains constant with the distance clearly favors the coexistence of various ionization parameters at each radius. This would imply a range in density and temperature at any given location, as already suggested, for example, by \citet{kin02} and \citet{ogle03}. Such a distribution may naturally arise as a result of thermal instabilities in outflows \citep{kk01}. In any case, the clumpy appearance of the [{O\,\textsc{iii}}] emission regions is very clear from the \textit{HST} images (see Fig. \ref{xray2oiiimaps} and \ref{xray2oiiimaps_2}), thus implying that our model in terms of a single cone of gas with density smoothly decreasing outwards is oversimplified. Again, the role of radio jets in confining different components may be crucial. Unfortunately, the limited \textit{Chandra} spatial resolution does not allow us to compare the substructures of the soft X-ray emission with those of the NLR and the radio jets, thus preventing us from drawing any firm conclusion on this issue.

\section{Conclusions}

We have presented a sample of eight Seyfert 2 galaxies included in the \citet{schm03} catalog of extended [{O\,\textsc{iii}}] images, with a \textit{Chandra} observation. All the sources but one present extended emission in soft X-rays, whose dimension and shape closely matches that of the [{O\,\textsc{iii}}]. The only exception, NGC~4507, does not show clear evidence for extended X-ray emission, but it is still consistent with the observed [{O\,\textsc{iii}}] emission.

The spectral analysis of the sources suggests that the most likely origin for the soft X-ray emission is in a gas photoionized by the nuclear continuum. The clearest piece of evidence comes from the 190 ks combined RGS spectrum of Mrk~3, which unambiguously appears to be produced in a photoionized gas with an important contribution from resonant scattering. In the other cases, where only spectra with CCD resolution are available, a `scattering' model (a powerlaw plus emission lines) is to be preferred to a `thermal' model either on statistical grounds or because of unphysical best fit parameters of the latter (quasi-zero abundances).

Therefore, we tested with the code \textsc{cloudy} if a solution in terms of a gas photoionized by the nuclear continuum which produces both the soft X-ray and the [{O\,\textsc{iii}}] emission is tenable. Such solutions exist and require the density to decrease with radius. In particular, the fact that the observed [{O\,\textsc{iii}}] to soft X-ray ratio is fairly constant up to large radii suggests that the density law should be close to $r^{-2}$, similarly to what found in the NLR of some objects.

In any case, these models are clearly oversimplified and do not take into account a number of details. In particular, there is evidence, both in the optical and in the X-rays, that various components are present at any radius, as also indicated by the clumpiness of the extended region when seen with the high spatial resolution of \textit{HST}. In this sense, the role of radio jets, almost always found to be very well correlated with the NLR/soft X-ray emission, is still under debate and could be very important at least for the morphology of the gas. However, the investigation of more complex models would require better data quality, which will unfortunately not be available for many years to come.

\acknowledgement
We would like to thank G. Matt for many useful discussions, D. Porquet for advice and the anonymous referee for valuable suggestions. We acknowledge D. Tapiador for support on the use of Grid Technologies (Globus and GridWay), J. Carter for help on reducing the RGS data and C. Gordon for the resolution of some issues with \textsc{Xspec} 12.2.0.

\bibliographystyle{aa}
\bibliography{sbs}

\end{document}